\documentclass[twocolumn]{aastex7}
\usepackage{amsmath}
\usepackage{float}
\usepackage{graphicx} 
\usepackage{booktabs}
\usepackage{multirow}
\usepackage{array}
\defcitealias{carter_benchmark_2024}{Carter \& May et al.}

\begin{document}

\title{Overcast mornings and clear evenings in hot Jupiter exoplanet atmospheres}
\author[0000-0002-3263-2251]{Guangwei Fu}\email{guangweifu@gmail.com}
\affiliation{Department of Physics and Astronomy, Johns Hopkins University, Baltimore, MD, USA}

\author[0000-0003-1622-1302]{Sagnick Mukherjee}\email{}
\affiliation{Department of Astronomy and Astrophysics, University of California, Santa Cruz, CA 95064, USA \\ }
\affiliation{Department of Physics and Astronomy, Johns Hopkins University, Baltimore, MD, USA \\ }

\author[0000-0002-7352-7941]{Kevin B. Stevenson}\email{}
\affiliation{JHU Applied Physics Laboratory, 11100 Johns Hopkins Rd, Laurel, MD 20723, USA}

\author[0000-0001-6050-7645]{David K. Sing}\email{}
\affiliation{Department of Physics and Astronomy, Johns Hopkins University, Baltimore, MD, USA}

\author[0000-0002-0786-7307]{Reza Ashtari}
\email{reza.ashtari@jhuapl.edu}
\affiliation{JHU Applied Physics Laboratory, 11100 Johns Hopkins Rd, Laurel, MD 20723, USA}

\author[0000-0001-6707-4563]{Nathan Mayne}\email{n.j.mayne@exeter.ac.uk}
\affiliation{Department of Physics and Astronomy, Faculty of Environment Science and Economy, University of Exeter, EX4 4QL, UK}

\author[0000-0003-3667-8633]{Joshua D. Lothringer}\email{}
\affiliation{Space Telescope Science Institute, Baltimore, MD}

\author[0000-0002-9705-0535]{Maria Zamyatina}\email{m.zamyatina@exeter.ac.uk}
\affiliation{Department of Physics and Astronomy, Faculty of Environment Science and Economy, University of Exeter, EX4 4QL, UK}

\author[0000-0001-8510-7365]{Stephen P. Schmidt}\email{sschmi42@jh.edu}
\altaffiliation{NSF Graduate Research Fellow}
\affiliation{Department of Physics and Astronomy, Johns Hopkins University, Baltimore, MD, USA}

\author[0000-0001-5097-9251]{Carlos Gascón}\email{}
\affiliation{Center for Astrophysics $\mid$ Harvard $\&$ Smithsonian, 60 Garden Street, Cambridge, MA 02138, USA}
\affiliation{Institut d'Estudis Espacials de Catalunya (IEEC), 08860 Castelldefels, Barcelona, Spain}

\author[0000-0002-0832-710X]{Natalie H. Allen}\email{}
\affiliation{Department of Physics and Astronomy, Johns Hopkins University, Baltimore, MD, USA}

\author[0000-0002-9030-0132]{Katherine A. Bennett}\email{}
\affiliation{Department of Earth and Planetary Sciences, Johns Hopkins University, Baltimore, MD, USA}

\author[0000-0003-3204-8183]{Mercedes L\'opez-Morales}\email{}
\affiliation{Space Telescope Science Institute, Baltimore, MD}
\affiliation{Center for Astrophysics $\mid$ Harvard $\&$ Smithsonian, 60 Garden Street, Cambridge, MA 02138, USA}

\begin{abstract}

Aerosols is an old topic in the young field of exoplanet atmospheres. Understanding what they are, how they form, and where they go has long provided a fertile playground for theorists. For observers, however, aerosols have been a multi-decade migraine, as their chronic presence hides atmospheric features. For hot Jupiters, the large day-night temperature contrast drives inhomogeneous thermal structures and aerosol distribution, leading to different limb properties probed by transit spectra. We present JWST NIRISS/SOSS spectra of morning and evening limbs for nine gas giants with equilibrium temperatures of $\sim$800–1700 K. By measuring feature size of the 1.4$\mu m$ water band for both limbs, we found three planets (WASP-39 b, WASP-94 Ab, and WASP-17 b) show prominent ($>$5$\sigma$) limb-limb atmospheric opacity difference with muted morning and clear evening limbs. The heavily muted water features on morning limbs indicate high-altitude (0.1 to 0.01 mbar) aerosols. To simultaneously have clear evening limbs requires processes with timescales ($\sim$day) comparable to advection to remove these lofted grains, and we found that both downwelling flow and dayside cloud evaporation could be plausible mechanisms. We hypothesize an empirical boundary—termed the “asymmetry horizon”—in temperature–gravity space that marks the transition where inhomogeneous aerosol coverage begins to emerge. Heterogeneous aerosol coverage is common among hot Jupiters. If unrecognized, limb averaging suppresses spectral features, mimicking high-mean-molecular-weight atmospheres, inflating inferred metallicity by up to 2 dex, and underestimating limb temperatures by as much as half. Finally, we introduce the Limb Spectroscopy Metric (LSM) to predict limb spectral feature size based on planet parameters.  

\end{abstract}


\nopagebreak
\section{Introduction}

In nature, symmetry provides elegance, and asymmetry leads to complexity. The transit light curve of an exoplanet is mostly symmetric, primarily determined by the relative size of two spherical objects, the planet and the star, shaped by gravity. Second-order effects, such as stellar heterogeneity, eccentricity, oblateness, atmospheres, natural satellites, and telescope systematics, bring in complexity and introduce asymmetry. 

One of the complex phenomena in exoplanet atmospheres is aerosols. The phase transition between gas and solid is a fundamental process in any planetary atmosphere. The presence of condensate clouds and photochemical hazes on exoplanets has been detected numerous times \citep{sing_hubble_2011, kreidberg_clouds_2014, brandeker_cheops_2022, fu_statistical_2017, grant_jwst-tst_2023, inglis_quartz_2024}, but our understanding of them has remained limited \citep{gao_aerosols_2021}. The large parameter space of aerosols physics \citep{helling_cloud_2022} and often multiple orders of magnitude of uncertainties on cloud microphysics models parameters \citep{powell_transit_2019} often lead to highly degenerate interpretations of the observations. This is compounded by uncertainties in the spatial transport of aerosols as predicted by atmospheric general circulation models (GCM) \citep{steinrueck_3d_2021}. 

The era of JWST has brought a leap in measured precision that enables us to detect ever smaller signals that arise from second-order effects. One such effect is atmospheric limb-limb asymmetry (\citealp{espinoza_inhomogeneous_2024, murphy_evidence_2024}; Sagnick Mukherjee et al. 2025 submitted). Limb-limb atmospheric differences in hot Jupiter atmospheres have been long predicted by GCMs \citep{showman_atmospheric_2009, parmentier_non-grey_2014, christie_impact_2021,fortney10}. The tidal locking induces day and night temperature differences that lead to eastward $\sim$km/s jets and a few hundred Kelvin temperature difference between morning and evening limbs. The temperature difference can then lead to differences in atmospheric scale height \citep{espinoza_inhomogeneous_2024} and chemistry \citep{kesseli_atomic_2022, savel_diagnosing_2023}. Another key resulting difference is the presence of aerosols, which is affected by both vertical and horizontal transport \citep{kempton_observational_2017}. A comparative study of cloud signatures on each limb for the same planet is a unique way to advance our understanding of cloud formation processes and spatially locate them in the atmosphere \citep{powell_transit_2019}. 

Here we present the morning and evening limb transit spectra for nine different hot Jupiters with equilibrium temperatures spanning 770 to 1740\,K and planetary surface gravities spanning 2.45 to 3.2 logg (cgs). All spectra use previously published archival JWST NIRISS/SOSS datasets. By uniformly analyzing and comparing them on a population level, we aim to search for first-order effects that govern the formation and dissipation of aerosols within hot Jupiter atmospheres. 

\begin{figure}[t]
    \centering
    \includegraphics[width=0.5\textwidth]{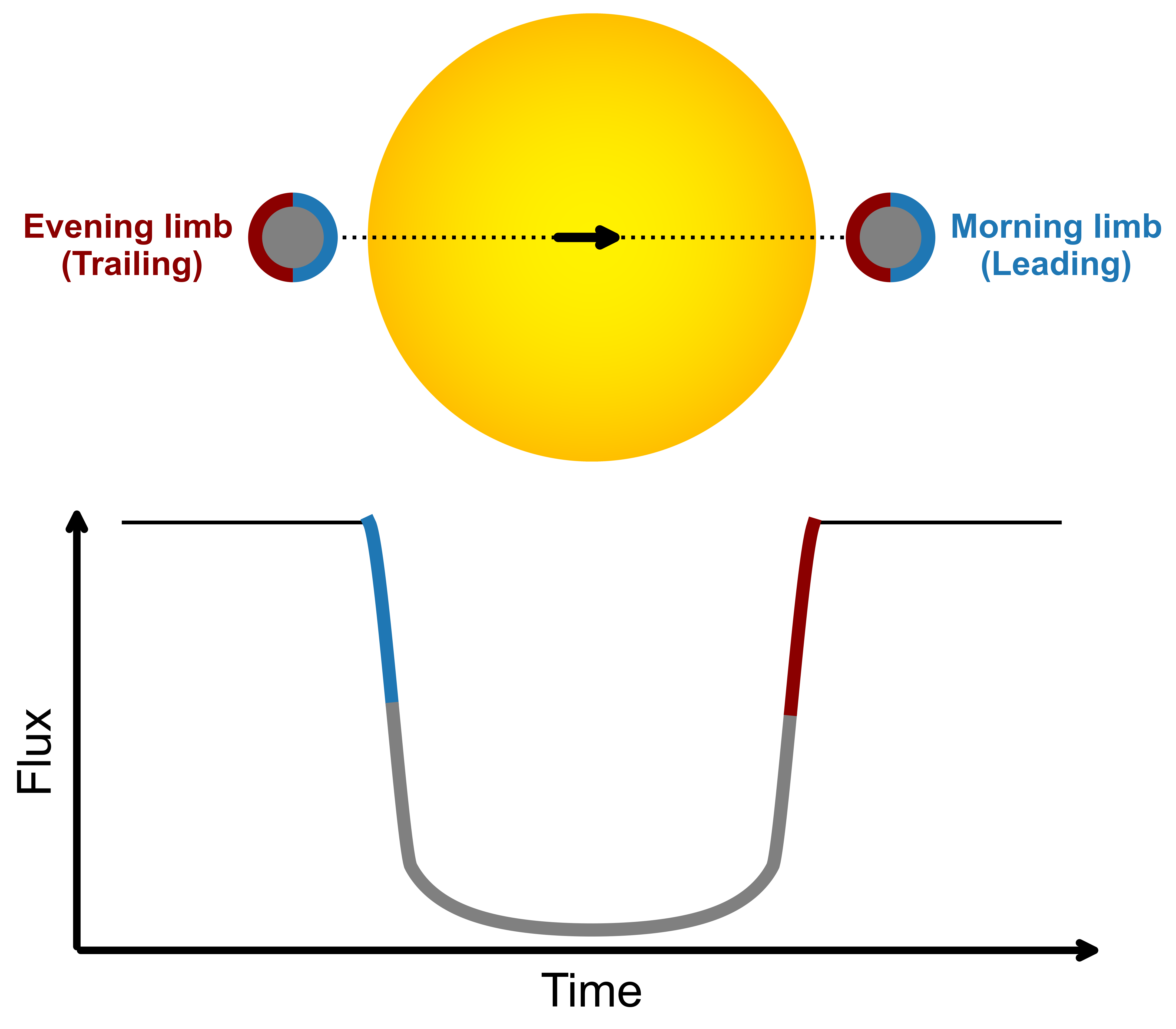}
    \caption{Diagram illustrating the transit geometry with corresponding morning (blue), evening (red), and combined (grey) limb signals in the transit light curve.}
\label{Fig1:cartoon}   
\end{figure}

\section{Methods}

\subsection{Definition of morning and evening limbs}

Hot Jupiters are typically assumed to be tidally locked due to their short orbital periods, and the atmospheric jet direction is expected to align with their orbital motion \citep{showman_equatorial_2011}. Based on these conventionally accepted assumptions, the leading limb, which enters during ingress, is expected to be cooler as it receives air from the planet’s nightside. In contrast, the trailing limb, which exits last during egress, is anticipated to be warmer due to air circulation from the hotter dayside. Based on this paradigm, the leading and trailing limbs are described as the morning and evening limbs, respectively (Figure \ref{Fig1:cartoon}) \citep{espinoza_inhomogeneous_2024, espinoza_constraining_2021}.

\subsection{Sample and instrument mode selection}

We collected all published JWST transmission spectra of giant exoplanets observed using NIRISS/SOSS. The focus on giant planets is due to the high signal-to-noise ratio (SNR) required to detect the subtle effects of limb-limb asymmetry. The number of light-curve points within the ingress and egress that contain the asymmetry information is roughly an order of magnitude less than the total number of points in transit; thus the limb spectral precision is approximately three times worse than the normal limb-averaged spectrum. This level of spectral quality degradation precludes our ability to meaningfully measure limb spectra for lower SNR targets, such as sub-Neptunes or planets around faint host stars. 

We focus on NIRISS/SOSS, which covers approximately 0.8–2.8 $\mu m$, due to its enhanced sensitivity to aerosols at shorter wavelengths, where gas to aerosol opacity ratio is relatively lower.

\begin{figure*}[]
    \centering
    \includegraphics[width=0.8\textwidth]{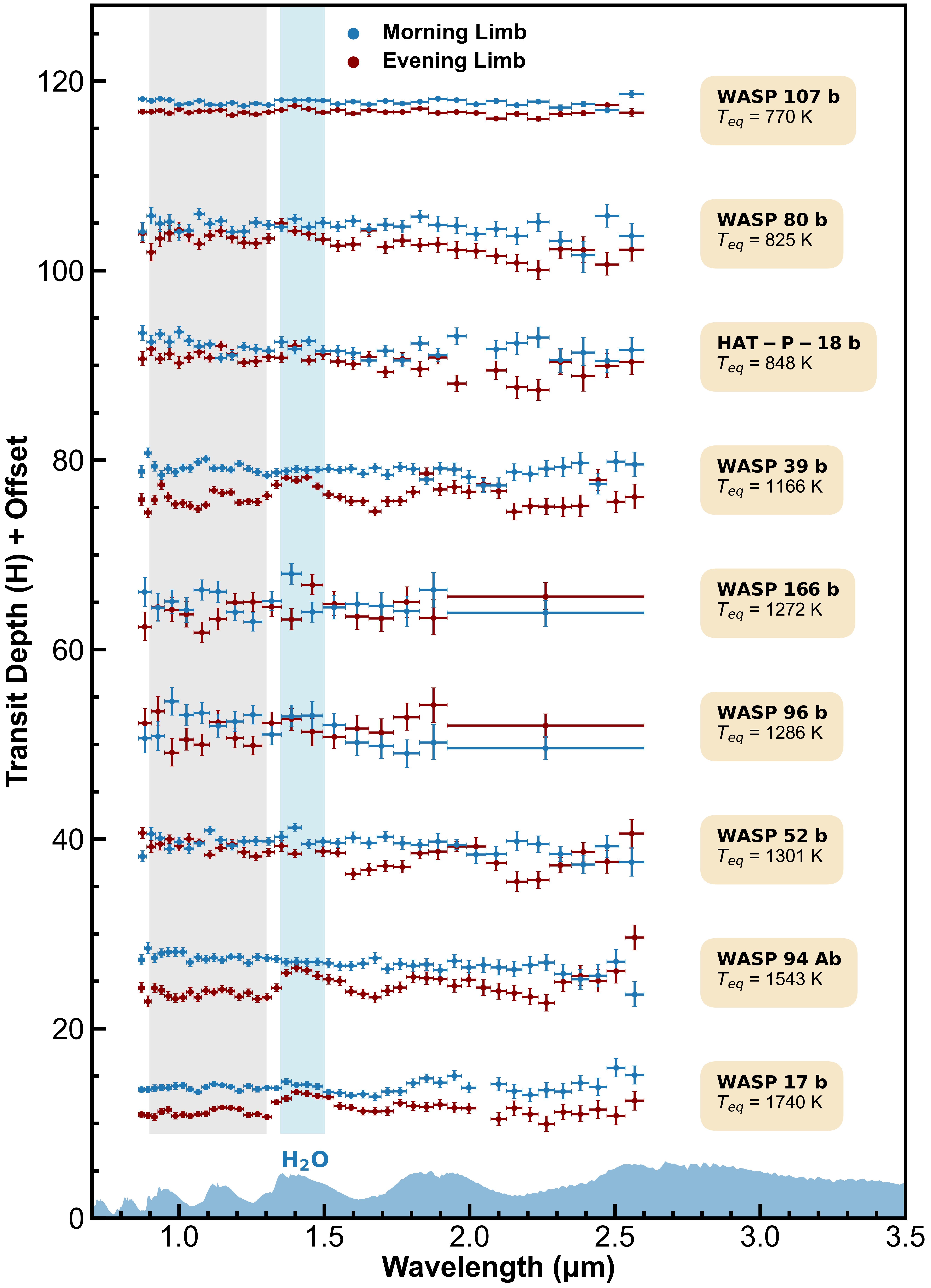}
    \caption{The morning (blue) and evening (red) limb spectra for the nine-planet sample used in this study. We normalize the spectra by their corresponding atmospheric scale heights (H) and apply a 13 H offset between planets, which are sorted by temperature. The water opacity is shown at the bottom. The vertical gray (0.9-1.3 $\mu$m) and light blue (1.35-1.5 $\mu$m) shaded areas represent the baseline and 1.4 $\mu$m water band regions, respectively. To eliminate spectral overlap between planets, the morning limb spectra are offset to be one scale height above the evening limb spectra based on the averaged inside of the water band transit depth values. We first measured the band average for both blue and gray regions for both limbs for each planet, and then calculated the relative differences between them (A$_H$).}
\label{Fig2:spectra}   
\end{figure*}

\subsubsection{Photometric bands selection}

Within the SOSS wavelength coverage, we select two photometric bands 0.9-1.3$\mu m$ and 1.35-1.5$\mu m$. The 1.4$\mu m$ water band has long been used to probe the presence of clouds using HST \citep{stevenson_quantifying_2016, fu_statistical_2017, brande_clouds_2024}. By comparing the relative absorption strength from inside this strong water band versus the outside baseline, we are able to quantify the strength of the aerosol muting effect, which can be indicative of the cloud deck pressure and particle size. 

The SOSS covers multiple water bands, including more around 1.9 and 2.5 $\mu m$. We choose to focus on the 1.4 $\mu m$ band for two reasons: (1) It has the highest SNR and thus is less vulnerable to systematic noises. (2) It is the most sensitive to aerosol muting effects.

\subsection{Data reduction}

\subsubsection{Limb spectra fitting}
The data reduction steps are broadly similar for each dataset as outlined in previous studies using SOSS \citep{fu_water_2022, davenport_toi-421_2025}. All datasets are downloaded from MAST starting from the \texttt{uncal.fits} files. We first use the default \texttt{jwst} pipeline to run stage1 to the \texttt{dark\_current} step. Then we performed group-level background subtraction on the \texttt{dark\_current.fits} file. Next, we processed the data through the \texttt{jump\_step} and \texttt{ramp\_fit\_step} to obtain the \texttt{rampfitstep.fits} files. Time and corresponding integration are then all saved for each observation. For each integration, we first clean out any leftover individual bad pixels by comparing them to the median image of the entire observation. Next, we find the first-order SOSS spectral trace by cross-correlating the SOSS psf for each column and find the maximum location in the vertical direction as a function of spectral dispersion in the horizontal direction. Then the spectrum is extracted using a 30-pixel wide aperture centered at the trace location. Each spectroscopic light curve is then cleaned for outliers temporally with a rolling median. The final cleaned light curves are then passed for the next steps of light curve fitting.

\begin{figure*}[]
    \centering
    \includegraphics[width=0.8\textwidth]{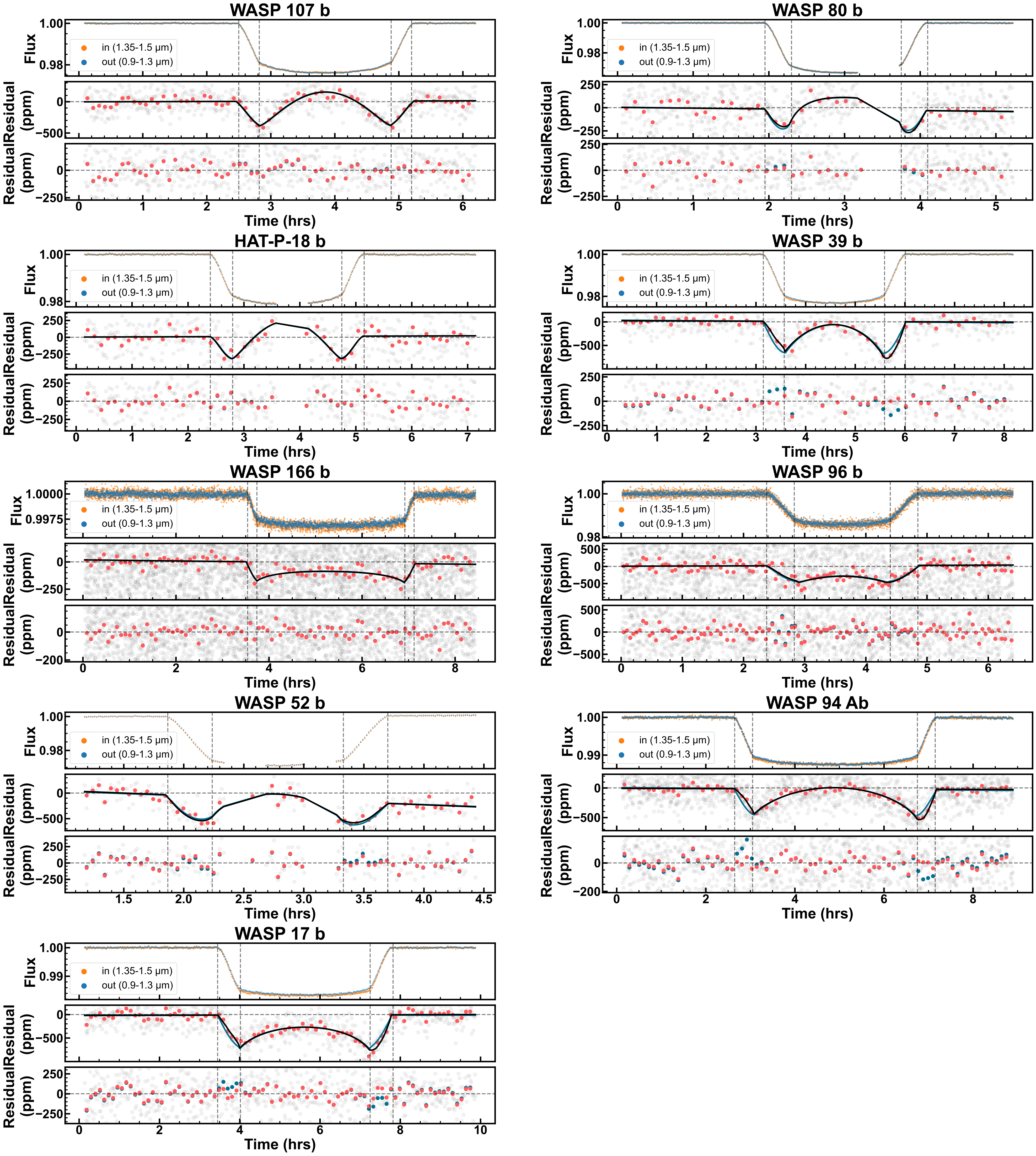}
    \caption{Transit light curves from inside the 1.4 $\mu$m water band versus the outside baseline are shown in the top panels for each of the nine planets. The differences between the two light curves are shown in the middle panels. This direct empirical comparison visualizes the wavelength-dependent asymmetric signals in the transit light curves. Residuals (Red) between two bands should be symmetric around ingress and egress if the planet has the same morning and evening limb spectral shapes regardless of limb darkening, mid-transit time, and eccentricity. The spot crossings are removed from the light curves. The best-fit one-circle (Teal) and two semi-circle (Black) transit models are overplotted in the middle panels. The mid-transit times are both fixed to the white-light mid-transit times. The respective residuals for both fits are shown in the bottom panels. The two semi-circle models can better fit ingress and egress for especially three planets, WASP 39 b, WASP 94 Ab, and WASP 17 b where we see strong morning and evening limb spectra difference in Figure \ref{Fig2:spectra}.}
\label{Fig3:lightcurve}   
\end{figure*}

The white light curve, which comes from summed spectroscopic light curves, is first used to fit common system parameters. We used \texttt{Catwoman} \citep{jones_catwoman_2020, espinoza_constraining_2021} combined with \texttt{emcee} \citep{foreman-mackey_emcee_2013} for the fitting. A total of nine free parameters are used, including visit-long slope, constant, mid-transit time, Rp, $\Delta Rp$, inclination, a/R$_{star}$, and two quadratic limb darkening parameters. The default \texttt{Catwoman} Rp1 and Rp2 parameters are reparameterized to Rp and $\Delta Rp$ through Rp1 = Rp - $\Delta Rp$/2 and Rp2 = Rp + $\Delta Rp$/2. The purpose of this reparameterization is to improve the mcmc sampling efficiency by alleviating the strong degeneracy between Rp1 and Rp2. Since we use \texttt{emcee}, an affine-invariant ensemble sampler, designed to effectively sample correlated parameter spaces, our reparameterization does not change the final results. However, this could be beneficial when using different parameter estimation methods.

The best fit white-light mid-transit time, inclination, and a/R$_{star}$ are then fixed for the spectroscopic channel fits. There are four free-fitting parameters including visit-long slope, constant, Rp, and $\Delta Rp$. The two quadratic limb darkening parameters are fixed to the \texttt{stagger} \citep{magic_stagger-grid_2015} 3D stellar grid values. The best fit spectroscopic Rp and $\Delta Rp$ are then converted back to rp1 (Evening) and rp2 (Morning). The transit depth, (Rp/Rs)$^2$, of each limb is then normalized by the corresponding atmospheric scale height of each planet \citep{fu_statistical_2024} and shown in Figure \ref{Fig2:spectra}. The planet parameters assumed in the scale height calculation are shown in Table \ref{tab:targets}, with a mean molecular weight of 2.3 amu and the same equilibrium temperature for each limb assumed for all planets. The limb water index (A$_H$) is calculated by measuring the mean transit depth between the in (0.9-1.3$\mu$m) and out (1.35-1.5$\mu$m) of the 1.4 $\mu$m water band in scale heights. The errors of the two bands are calculated by propagating the uncertainties of individual points, and then added in quadrature to get the total error for A$_H$.

\begin{figure}[t]
    \centering
    \includegraphics[width=0.45\textwidth]{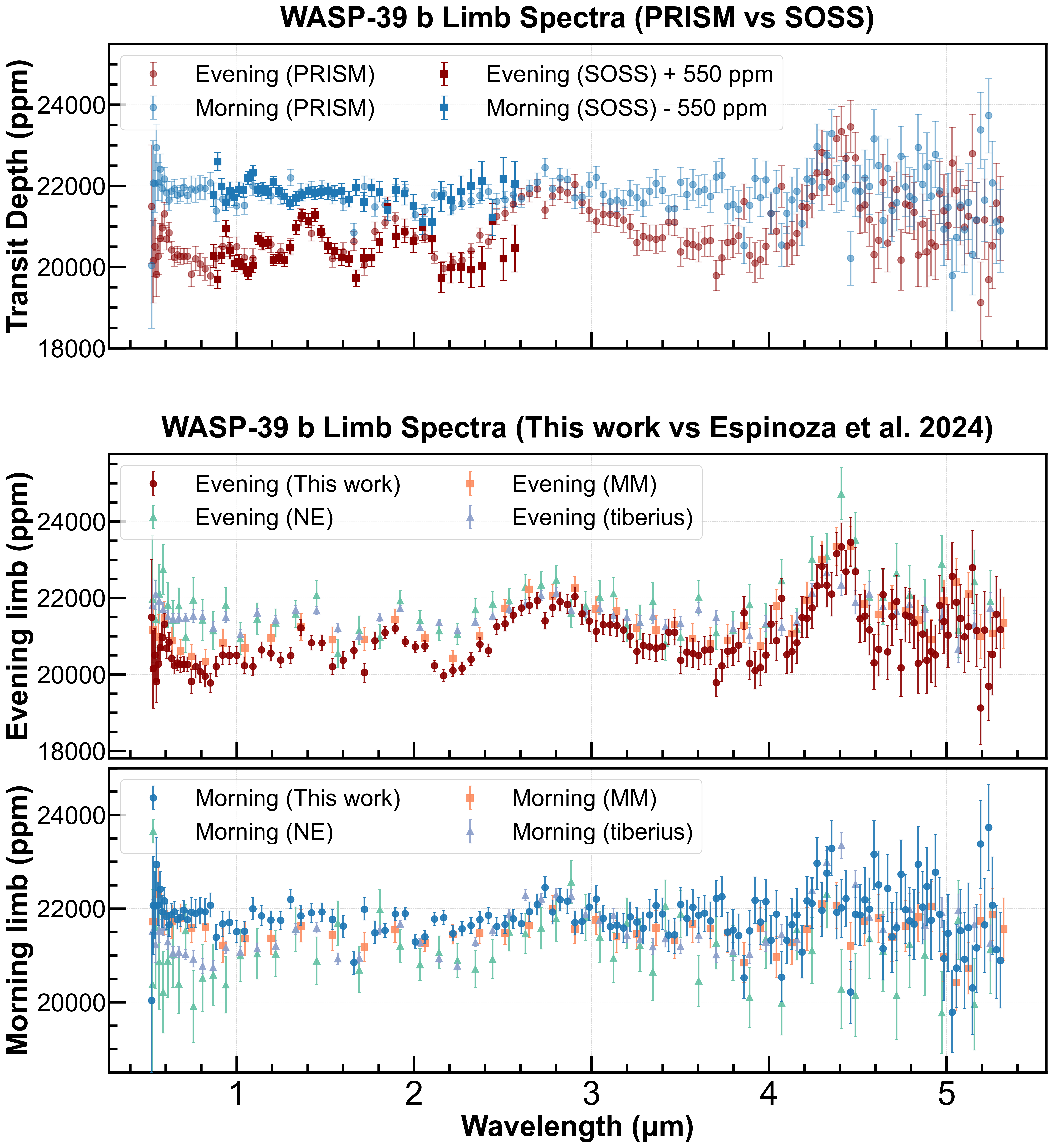}
    \caption{Comparison between morning and evening spectra for WASP-39 b using NIRISS/SOSS (This work) versus NIRSpec PRISM (This work) (Top panel). Comparison between NIRSpec PRISM (This work) evening (middle panel) and morning (bottom panel) limbs spectra to three reductions in Extended Figure 1 from \cite{espinoza_inhomogeneous_2024}, where the labels (NE, MM, and tiberius) are kept consistent from \cite{espinoza_inhomogeneous_2024} representing the three reductions. Beyond the constant vertical offsets introduced by T$_0$ uncertainties, both NIRISS/SOSS and NIRSpec PRISM show consistent wavelength-dependent spectral shapes with muted morning limbs and clear evening limbs. This shows that the observed limb-limb spectral shape difference is not introduced by any SOSS instrument-specific systematics.}
\label{Fig4:W39b_compare}   
\end{figure}

\subsubsection{Empirical visualization of limb-limb asymmetry}

The limb spectra are products of light curves fitting using a combination of instrument systematics and \texttt{Catwoman} model, which approximates the planet as two semicircles with different radii. While those assumptions are reasonable \citep{espinoza_inhomogeneous_2024}, it is difficult to visualize the exact source of the limb asymmetry signals from the light curves. To better illustrate the asymmetric nature of light curves and independently check the robustness of the light curve fits, we overplotted and measured the residuals for the binned spectroscopic light curves in and out of the 1.4$\mu$m water bands for all nine planets (Figure \ref{Fig3:lightcurve}). This simple light curve comparison between the two bands bypasses all the degenerate wavelength-independent orbital parameters (eccentricity and mid-transit time) and focuses on the wavelength-dependent asymmetry signals. 

The dominant features in the residuals are the different transit depths and limb-darkening-induced ingress and egress shapes from the two bands. However, the residuals from both these differences are expected to be symmetric around the mid-transit time. Therefore, any asymmetric residual will be caused by atmospheric opacity difference in the morning and evening limbs between these two wavelength bands. We can see in the three planets (WASP 39 b, WASP 94 Ab, and WASP 17 b) where morning and evening spectra differ the most (Figure \ref{Fig3:lightcurve}), their egress residuals are significantly lower than their ingress residuals with asymmetric shapes beyond the level of photometric scatter in the baseline. Among the other five planets, the residuals are more symmetric relative to the photometric scatter levels. This is consistent with the limb spectra shown in Figure \ref{Fig2:spectra}.

Asymmetric stellar heterogeneity can produce light curve asymmetry on the limbs via hotspots or faculae. However, such asymmetry would be randomly distributed between ingress and egress, where the data preferentially show lower egress than ingress residuals (Figure \ref{Fig3:lightcurve}). Additionally, stellar heterogeneity will not introduce clear water features in the limb spectra (Figure \ref{Fig2:spectra}) since the FGK host stars in the sample are too hot to have water present in their photosphere.

\begin{figure}[htp]
    \centering
    \includegraphics[width=0.45\textwidth]{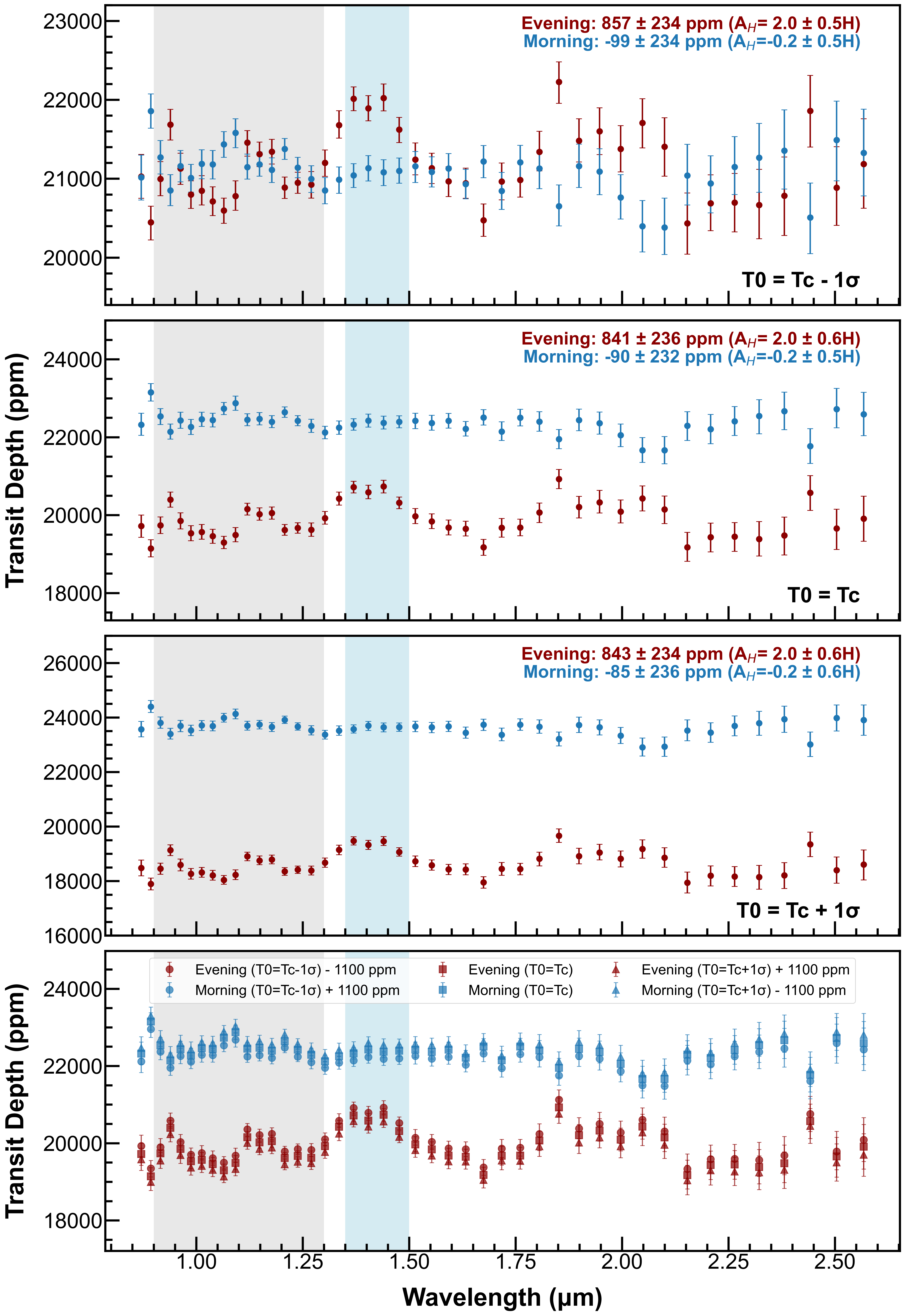}
    \caption{The absolute transit depth for the morning and evening limbs is sensitive to the exact T$_0$ value used for the spectroscopic light curve fits. However, the relative shapes of the spectra remain consistent. We demonstrated this with the WASP 39 b SOSS dataset by varying the T$_0$ by one $\sigma$ before and after the best-fit white light T$_0$. The corresponding A$_H$ value is insensitive to the T$_0$ variation.}
\label{Fig5:T0}   
\end{figure}

\subsubsection{Comparing NIRISS SOSS with NIRSpec PRISM}

To ensure the limb spectra differences are not introduced by SOSS-specific instrument systematics, such as detector readout properties, we also analyzed the NIRSpec PRISM dataset from \citetalias{carter_benchmark_2024} for WASP 39 b. We downloaded the spectroscopic light curves from Zenodo with bins\_scale1 spectral resolution. The light curve fitting steps are the same as described previously for SOSS. The morning and evening limb spectra are shown in Figure \ref{Fig4:W39b_compare}. Compared to the WASP 39 b SOSS spectrum, the PRISM spectrum shows consistent water features in the evening limb with more muted features in the morning limb, confirming the limb spectral shape differences to be instrument-independent.

\begin{figure*}[htp]
    \centering
    \includegraphics[width=0.8\textwidth]{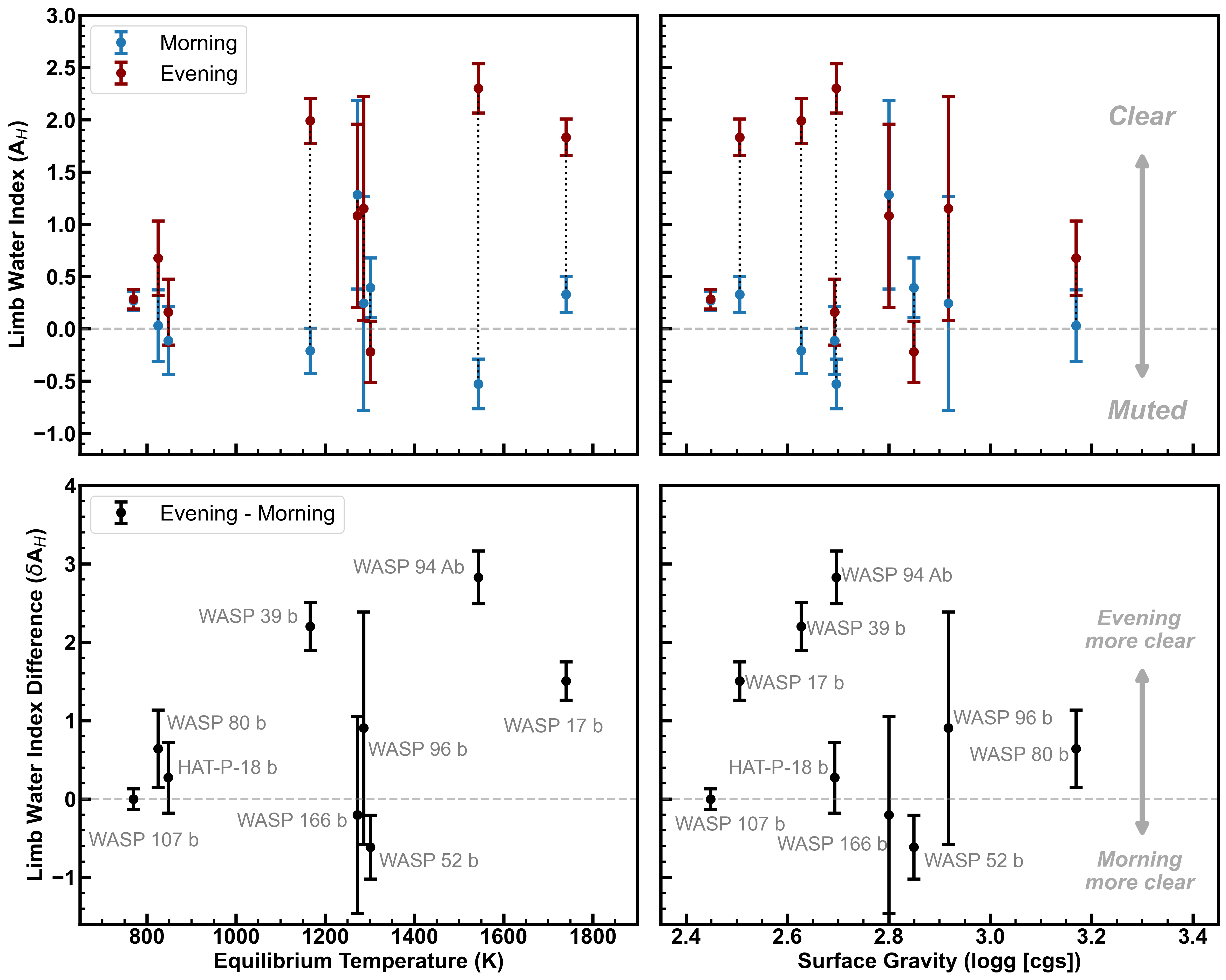}
    \caption{The top row shows the measured A$_H$ for the morning and evening limbs for each planet versus planet equilibrium temperature (left) and surface gravity (right). The bottom row shows the evening-morning difference ($\delta$A$_H$) for each planet. Neither planet temperature nor gravity appears to exhibit a simple relationship with A$_H$ or $\delta$A$_H$.}
\label{Fig6:Trend}   
\end{figure*} 

We also compared our NIRSpec PRISM limb spectra to the different reductions from Extended Data Fig. 1 in \cite{espinoza_inhomogeneous_2024}, which used the light curves from \cite{rustamkulov_early_2023}. Both limbs show similar features compared to the three reductions (NE, MM, and Tiberius). Our error bars are significantly smaller than those of NE, but are close to Tiberius and MM. This is likely due to the different choices of free or fixed fitting parameters and the spectroscopic light curves used. As described in the methods section of \cite{espinoza_inhomogeneous_2024}, the MM fit used the same four free parameters as in this work, and the NE fit used two more limb darkening coefficients and one jitter term.

\begin{figure}[htp]
    \centering
    \includegraphics[width=0.48\textwidth]{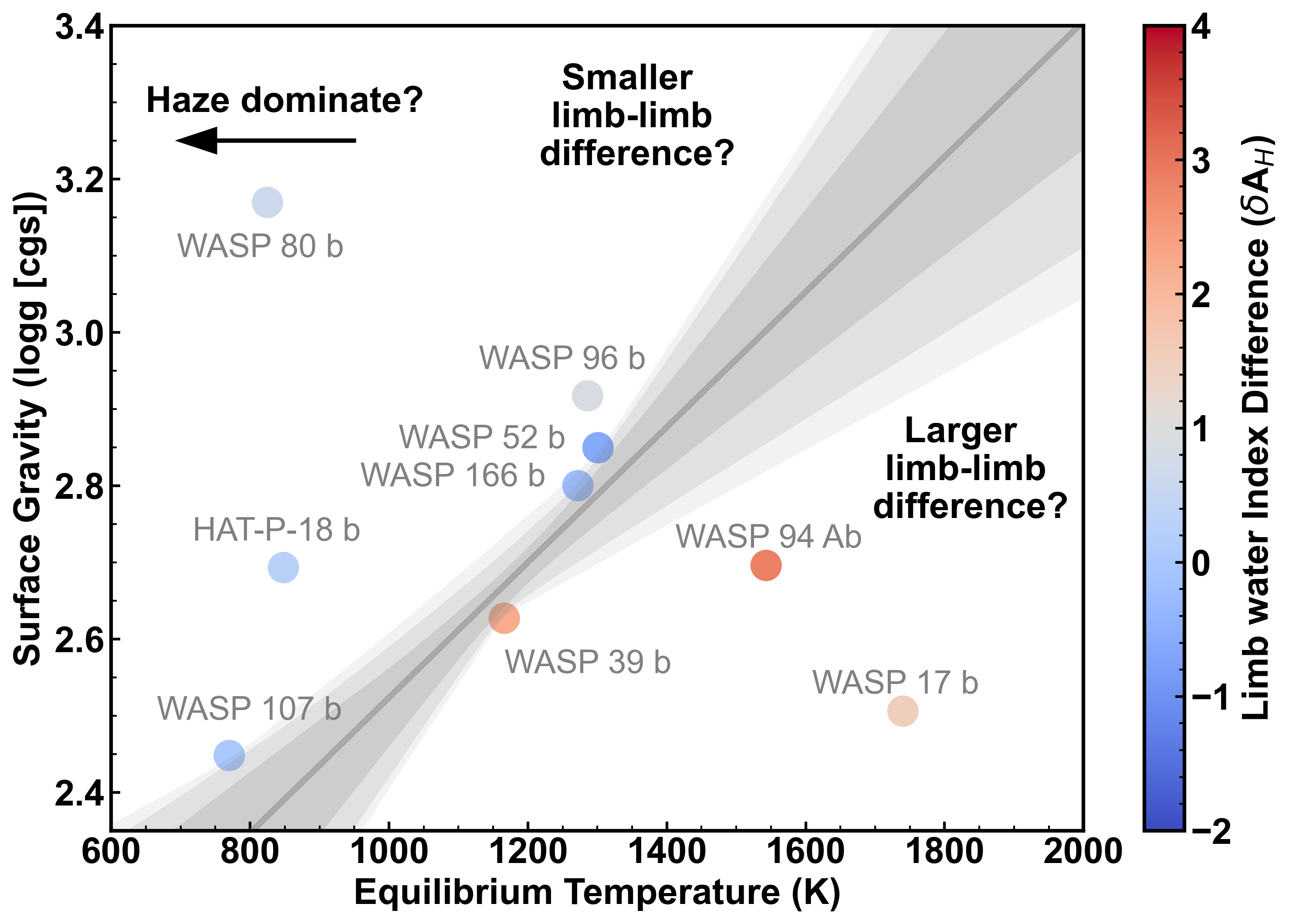}
    \caption{The measured $\delta$A$_H$ for each planet in planet equilibrium temperature versus surface gravity space. Planets with high $\delta$A$_H$ appear to cluster around the bottom right of the parameter space. We propose the asymmetry horizon (grey line), which outlines the transition region where aerosol-induced limb-limb asymmetry is expected to appear in the hot Jupiter atmosphere population. The line is empirically derived by fitting a 2D sigmoid function to the data and represents the contour where the function reaches half of its maximum value. The shaded gray area represents the 1, 2, and 3 $\sigma$ regions of uncertainties.}
\label{Fig7:asymmetry-horizon}   
\end{figure}

\subsubsection{Mid-transit time uncertainties vs. limb spectra}

Given the finite time sampling (typically $\sim$10-20 seconds using SOSS) resolution and photometric precision of observed light curves, the mid-transit time (T$_0$) uncertainties are typically on the order of $\sim$10s seconds. This level of precision is sufficient for measuring a typical limb-averaged transit spectrum, which is mostly constrained by the bottom of the light curve when the planet is entirely in the transit (T$_0$-insensitive). However, as the limb spectral signals are mostly determined by the points during ingress and egress, the absolute transit depth of each limb spectra becomes highly sensitive to the exact T$_0$ used. Fortunately, the relative shapes and differences between bands between morning and evening spectra remain insensitive to T$_0$ values. We demonstrated this by changing the T$_0$ value from the best fit white-light value to plus and minus one sigma and then fixing it for the spectroscopic channel fits (Figure \ref{Fig5:T0}). Then we measured the relative difference in transit depth between in and out of the water bands for both limbs. The consistent relative values show that changing T$_0$ only translates into a wavelength-independent offset between the morning and evening limb spectra.

\subsection{Empirical trends}

Relative measurements are always easier to make than absolute measurements in astrophysics. To bypass the absolute transit depth uncertainties induced by T$_0$, we measure the relative in (1.35-1.5 $\mu m$) versus out (0.9-1.3 $\mu m$) of water band transit depth for morning and evening limbs for all nine planets, with each being normalized by their corresponding atmospheric scale heights (A$_H$). As water vapor is expected to be present in all of these hot Jupiter atmospheres, lower A$_H$ would imply the spectra are muted by aerosols, and higher A$_H$ would indicate more prominent water features and clearer atmospheres. In addition, we calculated the difference of A$_H$ between limbs, $\delta$A$_H$, where we have subtracted the morning limb from the evening limb. $\delta$A$_H = 0$ means a similar spectral shape between the two limbs, positive $\delta$A$_H$ would indicate a relatively clear evening and muted morning limb, and negative $\delta$A$_H$ would represent the opposite.

We plotted A$_H$ and $\delta$A$_H$ against both planetary equilibrium temperature and surface gravity to search for any empirical trends (Figure \ref{Fig6:Trend}). While there are no clear linear correlations, there are a couple of notable features. (1) The morning limbs are generally more muted than the evening limbs. (2) The tentative double peak in $\delta$A$_H$ at WASP 39 b and WASP 94 Ab, with a valley around WASP 52 b.

\subsubsection{Asymmetry horizon}

We also plotted planetary equilibrium temperature versus surface gravity with $\delta$A$_H$ values color-coded (Figure \ref{Fig7:asymmetry-horizon}). All three planets that show prominent limb-limb spectral differences cluster around the bottom right part of the figure in the parameter space, meaning inhomogeneous aerosol-induced limb-limb asymmetry could be a function of both equilibrium temperature and surface gravity. To further quantify this transition empirically, we fitted a 2-dimensional sigmoid function to the data:

\begin{equation}
\delta A_H = \frac{z}{1 + e^{-k \left(m \cdot T_{eq} - logg + c\right)}}
\end{equation}

T$_{eq}$ is the equilibrium temperature in Kelvin, and logg is the log$_{10}$ of planet surface gravity in cm/s$^2$. We fitted the other four parameters (m, c, z, and k) using \texttt{emcee} \citep{foreman-mackey_emcee_2013} with the following median and 1$\sigma$ uncertainty values: 
$m = 0.0009^{+0.0003}_{-0.0002}$,\ 
$c = 1.6417^{+0.2381}_{-0.3415}$,\ 
$z = 2.0337^{+0.1647}_{-0.1665}$,\ 
$k = 579.21^{+290.56}_{-307.23}$.
Based on the fitted values, we plotted the contour line where the function reaches half of its maximum value in Figure \ref{Fig7:asymmetry-horizon}.

We propose an empirically derived "asymmetry horizon" in the T$_{eq}$ and logg parameter space. This line marks the transition where hot Jupiter atmospheres are expected to shift from homogeneous aerosol coverage to having clearer evening limbs than morning limbs.

\begin{figure*}[!ht]
    \centering
    \includegraphics[width=0.8\textwidth]{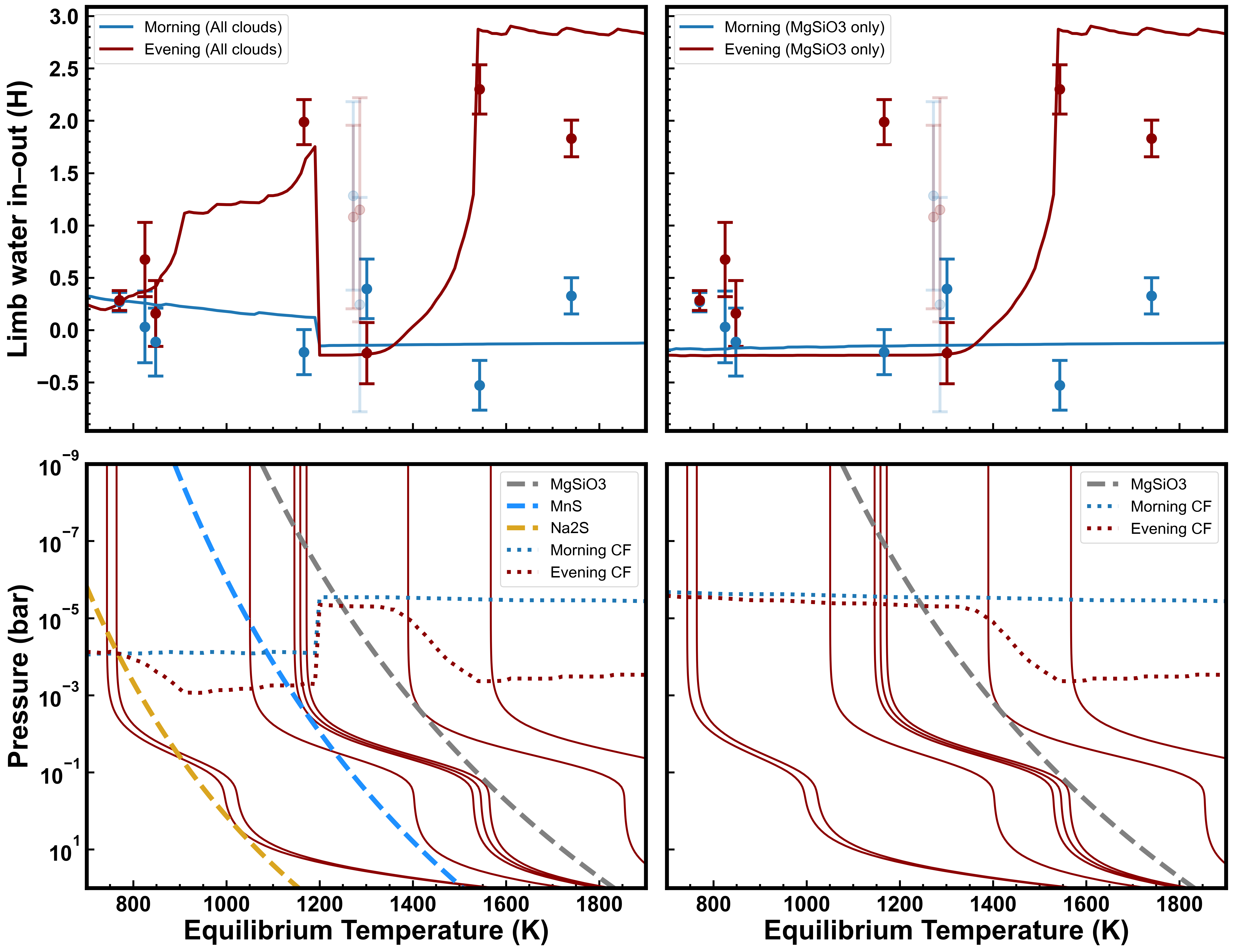}
    \caption{The top row shows the measured A$_H$ for the morning and evening limbs for each planet versus planet equilibrium temperature with best-fit \texttt{picaso} model tracks overplotted (See Figure \ref{Fig14:appendix_model_grid} for full spectra). The top left includes Na$_2$S, MnS, and MgSiO$_3$ clouds, and the top right only includes MgSiO$_3$ clouds. The bottom row shows temperature-pressure (TP) profiles for the evening limb for each planet, the morning limb TP profiles have the same parameterization except T$_{eq}$ is 350K lower. The cloud condensation lines are overplotted in dashed lines. The water band averaged peak contribution function pressure levels are shown in dotted lines for both limbs. }
\label{Fig8:1D_grid}   
\end{figure*}

\subsection{1D \texttt{PICASO} model grid}

To better interpret the observed empirical trends, we used the publicly available \texttt{PICASO} exoplanet atmospheric modeling code \citep{batalha_exoplanet_2019,mukherjee_picaso_2023} and the \texttt{VIRGA} cloud modeling code \citep{rooney_virga_2022, ackerman_precipitating_2001}. We constructed a parametric 1.5D atmospheric model to simultaneously generate transmission spectra for the morning and evening limbs. To model the $T(P)$ profile, we used the Guillot parametrization \citep{guillot_radiative_2010} for each planet limb. We assumed the same values of $\log(\gamma)$=-0.5, $\log(\kappa_{IR})$=-1.5, and $T_{\rm int}$=200 K for generating $T(P)$ profiles. The same value of $T_{\rm int}$ ensured that the $T(P)$ profiles for both planet limbs converge together at deeper pressures, even though they may differ significantly in the upper atmosphere. We let a difference of $\delta(T_{\rm eq})$ between the $T(P)$ profiles of the limbs to account for the limb-to-limb temperature difference. We assume thermochemical equilibrium for calculating the chemical abundance of each planet limb. 

We use \texttt{VIRGA} to simulate the clouds in each planet limb, with the \texttt{PICASO} $T(P)$ profiles as inputs. We include three common cloud species in our calculations-- Na$_2$S, MnS, and MgSiO$_3$. There are many more proposed cloud species for hot Jupiter atmospheres and an extensive exploration with model grids is beyond the scope of this study. We use optical constants from \citet{huffman67,khachai09,montaner79,scott96} for these minerals. The mass mixing ratios of these cloud species are fixed to solar composition values. The vertical distribution of clouds and the cloud particle sizes are controlled by two parameters in the \texttt{VIRGA} cloud model-- the cloud sedimentation efficiency ($f_{\rm sed}$) and the eddy diffusion coefficient $K_{\rm zz}$. The $f_{\rm sed}$ parameter is a ratio of the rate of sedimentation of cloud particles and their convective velocity. A smaller $f_{\rm sed}$ produces vertically lofted cloud decks, while a higher $f_{\rm sed}$ produces more vertically compact cloud decks. The $K_{\rm zz}$ parameter influences both the cloud particle sizes and the vertical extent of clouds significantly. A higher $K_{\rm zz}$ value allows bigger cloud particles to remain aloft. In our models, we fix the $K_{\rm zz}$ to 10$^{10}$ cm$^2$/s, and as a result, the vertical cloud distribution profile is determined by the $f_{\rm sed}$ parameter.

With this model setup, we generate a grid of transmission spectra for a generic planet (R$_p$=1R$_J$, M$_p$=0.3M$_J$) around a Sun-like star ($T_{\rm eff}$=5500K, R$_s$=1R$_{solar}$) at a spectral resolution of R$\sim$15,000. All model transit spectra are normalized by the atmospheric scale height, assuming the same equilibrium temperatures for both limbs. This allows us to compare the model grid to the measured limb water indices (A$_H$) across our nine-planet sample with the same normalization. The following model parameters are fixed: atmospheric metallicity (3x solar), C/O (solar), T$_{int}$ (200K), TP profile uses a Guillot profile ($\log(\gamma)$=-0.5 and $\log(\kappa_{IR})$=-1.5) \citep{guillot_radiative_2010}, evening limb temperature (T$_{eq}$), the vertical eddy diffusion coefficient ($K_{\rm zz}=10^{10}$ cm$^2$/s), and the included cloud species (Na$_2$S, MnS, and MgSiO$_3$). Since the limb temperature difference $\delta(T_{\rm eq})$ is expected to increase with T$_{eq}$, we parameterized it as a linear function of T$_{eq}$ ($\delta(T_{\rm eq})$=slope$\times$(T$_{eq}$-500K)) with the minimal value set to 0K when T$_{eq}$ is less than 500K. We then grid search through the following parameters: slope between T$_{eq}$ and $\delta(T_{\rm eq})$, $\log_{10}$($f_{\rm sed}$) and temperature to turn off MgSiO$_3$ condensation ($T_{\rm off}$). 

We included the T$_{off}$ parameter to check whether the limb-asymmetry trends somewhat reflect the L/T transition observed in brown dwarfs, where the photosphere very rapidly loses clouds near $T_{\rm eff}{\sim}1300$ K within a very small $T_{\rm eff}$ window of $\sim$100-200 K. At $T_{\rm eq}$ below $T_{\rm off}$, we remove all MgSiO$_3$ condensation from our model atmosphere. Through the grid search, we found the best-fitting parameters are slope=0.66, $\log_{10}$($f_{\rm sed}$)=-2.6, and $T_{\rm off}$=1200 K. The morning limb temperature is then determined as T$_{morning}$=T$_{eq}$ - 0.66$\times$(T$_{eq}$-500K). The best-fitting model tracks are plotted in the top left panel of Figure \ref{Fig8:1D_grid}. The full spectra of the track are shown in Figure \ref{Fig14:appendix_model_grid}. We then only included the MgSiO$_3$ cloud for all temperatures using the same best fit $\delta(T_{\rm eq})$ and $\log_{10}$($f_{\rm sed}$) to generate the top right panel tracks. The corresponding $T(P)$ profiles of the evening limb with peak averaged water band (1.35-1.5 $\mu$m) contribution function and cloud condensation lines are shown in the bottom two panels.

\section{Discussion}

\subsection{Clear evenings and muted mornings}

Based on the limb water index (A$_H$), the morning limbs are generally muted relative to the evening limbs (Figure \ref{Fig6:Trend}). All morning limbs have A$_H$ within 2 $\sigma$ of zero, indicating predominantly featureless spectra from 0.9 - 1.5 $\mu m$. The evening limbs are much less uniform, with WASP 39 b, WASP 94 Ab, and WASP 17 b showing strong water features, while the other five planets do not. The relative Evening-Morning difference ($\delta$A$_H$) shows that WASP 39 b, WASP 94 Ab, and WASP 17 b all have relatively clear evening limbs and overcast morning limbs. WASP 107 b, WASP 80 b, HAT-P-18 b, and WASP 52 b have relatively muted water features on both limbs shortward of $\sim$1.5$\mu m$. WASP 96 b and WASP 166 b are inconclusive due to their large error bars. Overall, evening limbs show stronger water molecular absorption features than morning limbs. 

\subsection{Temperature difference between limbs}

The HJ day-night temperature differences introduces non-uniform limb temperatures, and morning limbs are expected to be colder than the evening limbs. This temperature gradient can lead to (1) chemical composition and (2) feature size differences between the limbs \citep{roth_hot_2024,arnold_out_2025,zamyatina_observability_2022,powell_transit_2019}. For (1), since water is expected to be the main absorber in the SOSS wavelength and have uniform abundance between limbs at this temperature range, A$_H$ is insensitive to chemical composition difference between the limbs. For (2), the hotter temperatures increases the absolute feature sizes on the evening limbs to the first order, and less affect the relative water feature amplitudes. As discussed previously, since the absolute transit depth for each limb is highly sensitive to the T$_0$, we can not reliably constrain the limb temperatures from the absolute transit depth. Fortunately, A$_H$ is a relative measurement of the water feature amplitude, and changing temperature can not remove or create water features. Our A$_H$ measurement is designed to circumvent the effects from temperature differences between limbs, and focuses on capturing the varying muting effects of inhomogeneous aerosols coverages.

\begin{figure}[htp]
    \centering
    \includegraphics[width=0.45\textwidth]{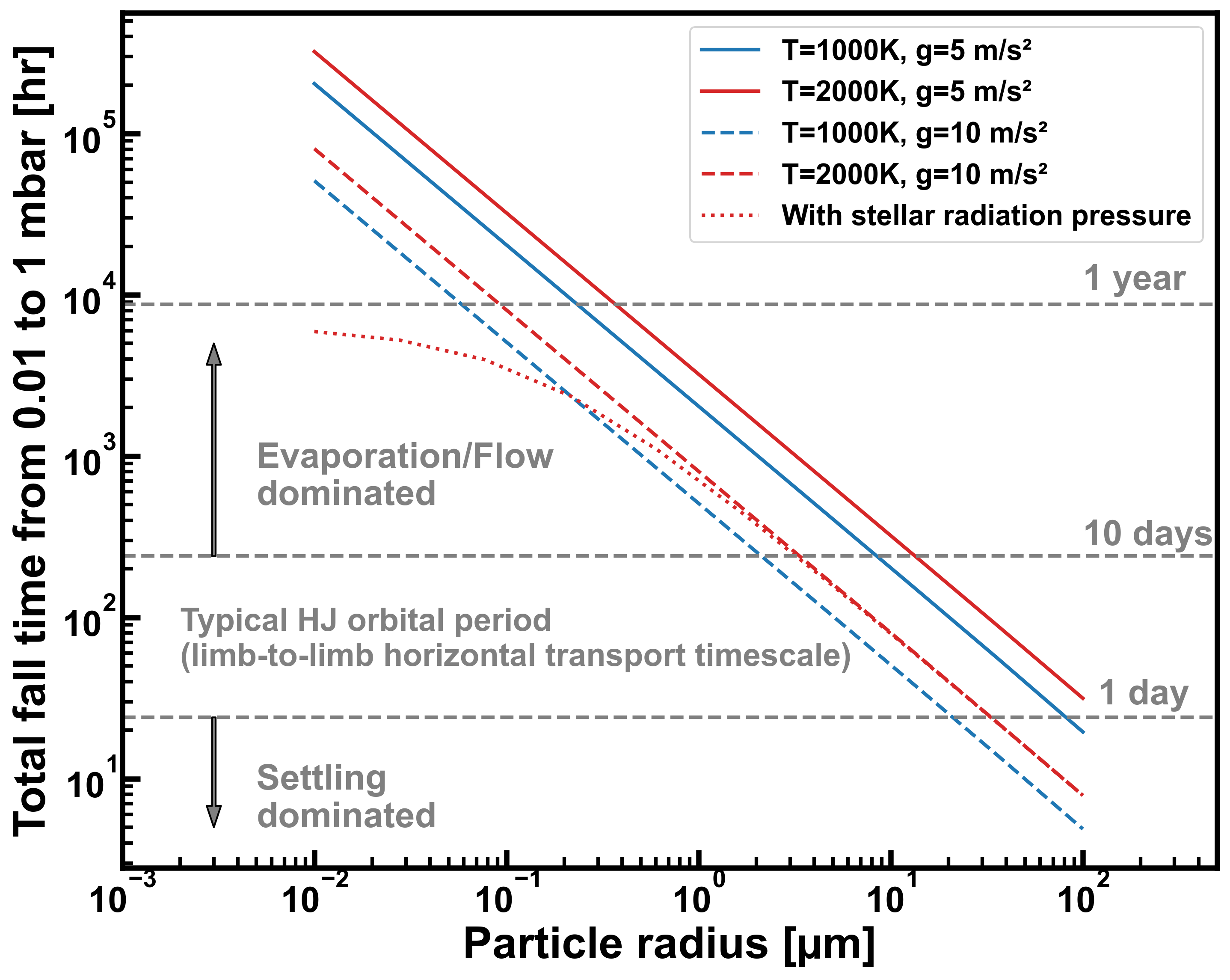}
    \caption{Calculated total fall time from 0.01 to 1 mbar as a function of particle size based on \cite{parmentier_3d_2013} for two temperatures and surface gravity. The effect of stellar radiation pressure is estimated by scaling up the gravity term based on $\beta$ as a function of grain size. To qualitatively match the observed muted morning and clear evening limbs, the total fall time should be comparable to or shorter than the advection timescale ($\sim$days). For large particles $(>\sim 10 \mu m)$, gravitational settling alone is sufficient. For small grains $(<\sim 1 \mu m)$, settling is not enough, additional mechanisms such as evaporation and vertical flow are needed.}
\label{Fig9:settle_timescale}   
\end{figure} 

\subsection{Possible mechanisms}

To generate the large morning and evening limb spectral shape differences on the order of a few atmospheric scale heights, aerosols need to be lofted to high altitudes ($\sim$1e-4 to 1e-6 bars) on the morning terminator and evaporate or settle to low altitudes ($\sim$1e-3 to 1e-2 bars) by the evening terminator. There are three main competing processes: gravitational settling, atmospheric flow, and condensation/evaporation cycles.

\subsubsection{Gravitational settling and stellar radiation pressure}

What goes up must come down. Gravity pulls on everything in the atmosphere, and the velocity can be approximated with the terminal fall speed. Assuming silicate grains in H$_2$-dominated atmospheres, we used the formula as outlined in \cite{parmentier_3d_2013} to calculate the total time it takes for a particle to fall from 0.01 to 1 mbars as a function of grain size. We calculated for four cases where the atmospheric temperatures are at 1000 and 2000 K with surface gravity at 5 and 10 m/s$^2$ (Fig \ref{Fig9:settle_timescale}). This is a representative range for planets in this study. For the gravitational settling process to be the main driver for muted morning and clear evening limbs, the particles need to be large ($\sim>$10$\mu m$). Small ($\sim<$1$\mu m$) particles are expected to stay lofted for much longer and would require atmospheric vertical flow or dayside evaporation to remove them.

Stellar radiation pressure could be another factor that accelerates the settling of grains in the dayside atmosphere (Owen $\&$ Murray-Clay 2025 submitted). It acts as an additional force to gravity that can push the grains downward towards deeper atmospheric layers. To the first order, the ratio between radiation pressure and gravity can be expressed as:

\begin{equation}
\beta \;=\; \frac{F_{\mathrm{rad}}}{F_g}
          \;=\;
          \frac{\displaystyle \frac{\pi r^{2} Q_{\mathrm{pr}} F_\star}{c}}
               {\displaystyle \frac{4}{3}\pi r^{3}\rho g}
          \;=\;
          \frac{3\,Q_{\mathrm{pr}}\,F_\star}{4\,r\,\rho\,c\,g}.
\end{equation}

The $r$ is particle radius, Q$_{pr}$ is the radiation‑pressure efficiency factor, F$_\star$ is the stellar flux, c is the speed of light, $\rho$ is grain density, and g is planet gravity. Assuming 0.5 micron sized silicate grains ($\rho$=3000~kg/m$^3$) in the atmosphere of a typical hot Jupiter (g=10~m/s$^{2}$) around a sun-like star (F$_\star$=1.5e6 Wm$^{-2}$ at 0.03 AU), the Q$_{pr}$ can be approximated to around 1 \citep{pawellek_dust_2019}. The calculated $\beta$ is 0.25, meaning radiation pressure can have comparable effects as gravity to around the same order of magnitude. 

\begin{figure}[htp]
    \centering
    \includegraphics[width=0.45\textwidth]{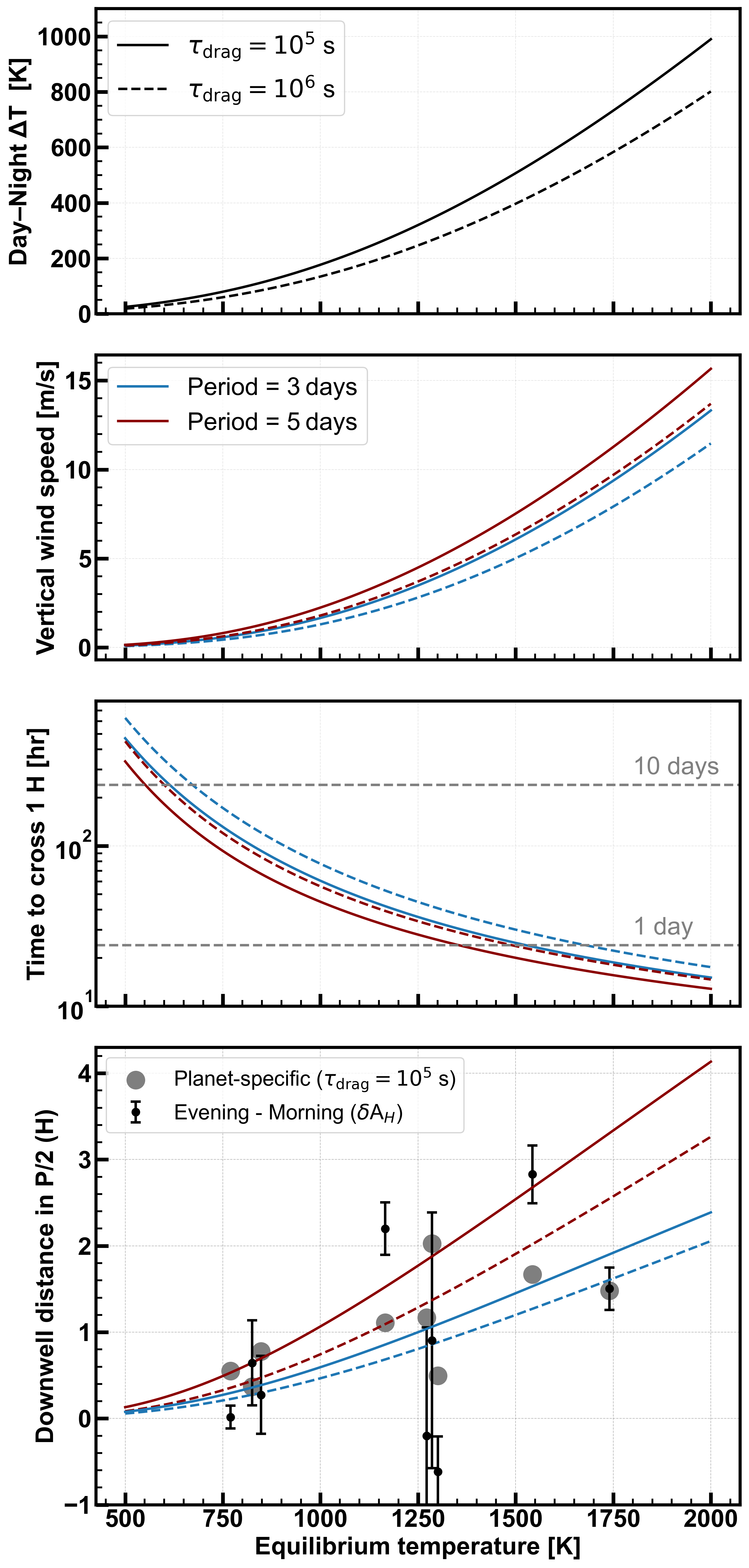}
    \caption{Day-night temperature difference is the main driver for large-scale flows on hot Jupiter atmospheres. We assumed two empirically motivated \citep{komacek_atmospheric_2017} day-night $\Delta$T versus T$_{eq}$ correlation (top panel) and then derived the expected vertical wind speeds (second panel) \citep{komacek_vertical_2019} and the time needed to cross 1 scale height (H) vertically (third panel). Then we calculated the number of H downwelling flows can travel in half of the planet's period (P/2) time (bottom panel) with planet-specific estimates (gray) and compared to measured $\delta$A$_H$ values. Downwelling flows can remove aerosols on a similar timescale as day-night flow, leading to clearer evening than morning limb spectra.}
\label{Fig10:vertical_timescale}   
\end{figure}

Based on the assumption above, we multiplied the gravity term by 1+$\beta$ in the terminal fall velocity calculations as a function of particle size. The modified total fall time from 0.01 to 1 mbars is shown in the dotted line for the T=2000 K and g=10 m/s$^2$ case (Figure \ref{Fig9:settle_timescale}). We assumed Q$_{pr}$=1 for all particle sizes. While this assumption is true for large particles ($>1 \mu m$), Q$_{pr}$ is expected to drop off for small particles ($<0.1\mu m$) where Rayleigh scattering is dominant. Our estimation here is therefore an upper limit on the effect of radiation pressure on small particles. While radiation pressure can shorten the time needed for grains to settle in the atmosphere, it mostly affects smaller grains ($\sim<0.1 \mu m$). The shortened timescales are still on the order of a year, much higher than the typical hot Jupiter advection timescale of days. For larger particles, this effect is negligible and gravity dominates.

\subsubsection{Atmospheric vertical flow}

The difference in received stellar flux between the day and night sides of hot Jupiter atmospheres drives the temperature gradient and large-scale flows. These flows can horizontally and vertically transport gas and aerosols \citep{showman_atmospheric_2009}. The vertical flows can transport aerosols upwards on the morning limb and downwards on the evening limb from the upwelling and downwelling flows at the terminators \citep{steinrueck_3d_2021}, qualitatively matching the observed muted morning and clear evening limb spectra (Figure \ref{Fig2:spectra}). To fully simulate the detailed atmospheric circulation with coupled cloud microphysics models is complex \citep{komacek_patchy_2022} and beyond the scope of this work, but we can make order-of-magnitude estimates on the vertical transport timescales for typical hot Jupiters.

\begin{table*}[t]
    \centering
    \begin{tabular}{c|c|c} 
        \hline\hline  
        \bfseries Planet & \bfseries Program ID & \bfseries Publications \\ 
        \hline 
        WASP 107 b   & GTO 1201 & \cite{murphy_panchromatic_2025}, \citep{krishnamurthy_continuous_2025} \\ 
        \hline 
        WASP 80 b   & GO 5924 & Gascon in prep \\ 
        \hline 
        HAT-P-18 b  & COM 2734 & \cite{fu_strong_2022}, \cite{fournier-tondreau_near-infrared_2023} \\ 
        \hline 
        WASP 39 b   & ERS 1366 & \cite{feinstein_early_2023} \\ 
        \hline 
        WASP 96 b   & COM 2734 & \cite{radica_awesome_2023}, \cite{taylor_awesome_2023} \\ 
        \hline 
        WASP 166 b  & GO 2062 & \cite{mayo_detection_2024} \\ 
        \hline 
        WASP 52 b   & GTO 1201 & \cite{fournier-tondreau_transmission_2024} \\ 
        \hline 
        WASP 94 Ab  & GO 5924 & \cite{mukherjee_cloudy_2025} \\ 
        \hline 
        WASP 17 b   & GTO 1353 & \cite{louie_jwst-tst_2024} \\ 
        \hline
    \end{tabular}
    \caption{JWST dataset information for each planet used for this study.}
    \label{tab:programID}
\end{table*}

Based on the day-night temperature difference, we can determine the characteristic horizontal wind speed and relate that to the expected vertical wind speed (See equations 9 to 13 from \cite{komacek_vertical_2019}) (Figure \ref{Fig10:vertical_timescale}). Then we can calculate the time it takes to travel across one atmospheric scale height as a function of planet equilibrium temperatures (Bottom panel Figure \ref{Fig10:vertical_timescale}). For the vertical flow to effectively loft and sink aerosols to produce muted morning and clear evening limbs, the vertical transport timescale needs to be comparable to the horizontal advection timescale ($\sim$day). As the planets increase in equilibrium temperatures, the vertical wind speeds are expected to increase, leading to shorter vertical transport timescales and more effective downwell flows that can sink aerosols. From vertical transport timescales, we can then calculate the number of scale heights downwelling flows can cross on the timescale of half the orbital period (P/2). This downwell distance in P/2 increases with larger day-night temperature gradients and longer orbital periods.

To the first order, measured $\delta$A$_H$ (lower panels of Figure \ref{Fig6:Trend}) traces the lower limits of how much aerosols need to travel vertically during the time it flows from the morning to evening limb. We can compare $\delta$A$_H$ to the predicted downwell distances based on each planet's equilibrium temperature, period, surface gravity, and radius (Figure \ref{Fig10:vertical_timescale} gray points). While the analytical prediction does not fully capture the variation in the data, it broadly agrees with the general trend of hotter planets showing higher $\delta$A$_H$.

\subsubsection{Condensation and evaporation}

Condensate clouds form and evaporate quickly ($\sim$mins) as the planet's temperature profiles cross the condensation curves \citep{parmentier_transitions_2016}. The large day-night temperature differences on hot Jupiters therefore provide the ideal environment for cloud cycling between nightside formation and dayside evaporation. To have a clear evening and cloudy morning, the clouds needs to form on the cooler planet's nightside and dissipate on the planet's dayside. To the first order, the planet's dayside and nightside temperatures need to straddle a cloud condensation line for this mechanism to take place. As the hotter evening limb moves right past the condensation line with clouds dissipating, the cooler morning limb still lags behind and stays on the left of the condensation line, continuously forming clouds. Since the cloud condensation line for a given cloud species is fixed in temperature and only weakly changes with metallicity \citep{visscher_atmospheric_2010}, only one peak of evening limb A$_H$ can be created with one cloud species. Indeed, when only MgSiO$_3$ cloud is included, we see one divergence between the morning and evening limb A$_H$ around 1600K (top right panel, Figure \ref{Fig8:1D_grid}).

However, one divergence could not explain the clear evening limb of WASP-39 b, indicating the need for more model complexity beyond just the condensation/evaporation cloud cycling of MgSiO$_3$. One way to better fit the clear evening limb of WASP 39 b is to create another divergence at cooler temperatures. To achieve that, we added Na$_2$S and MnS clouds (top left panel, Figure \ref{Fig8:1D_grid}). This shows that when assuming cloud condensation/evaporation as the main mechanism for observed limb spectra, the data support the presence of at least two distinct cloud populations in hot Jupiter atmospheres. There are many other dimensions to increase aerosol modeling complexity that could potentially explain the data \citep{kiefer_fully_2024, zamyatina_observability_2022, steinrueck_3d_2021}, and a comprehensive exploration of different and more complex modeling approaches is beyond the scope of this paper.

Besides adding a second population of clouds (Na$_2$S and MnS) at cooler temperatures, MgSiO$_3$ cloud needs to dissipate between 1200 K and 1400 K. Otherwise, all models cooler than the condensation temperature of MgSiO$_3$ will remain cloudy \citep{parmentier_transitions_2016}. In the previously mentioned 1D grid model, this dissipation process is determined by T$_{off}$ where the MgSiO$_3$ cloud opacity is turned off. The exact dissipation mechanism of MgSiO$_3$ clouds below $\sim$1300 K is currently unknown with possible explanations including nightside rainout \citep{komacek_patchy_2022} and horizontal transport \citep{bell_nightside_2024, komacek_atmospheric_2017, lewis_temperature_2022}. This seemly sudden clearing up of the atmosphere around 1300 K has also been observed in the L/T transition of the brown dwarfs \citep{suarez_ultracool_2022}. Processes including sedimentation \citep{stephens_08145_2009}, patchy clouds \citep{marley_patchy_2010}, vertical mixing \citep{ackerman_precipitating_2001}, and thermo-chemical instability \citep{tremblin_cloudless_2016} have been proposed to explain the L/T transition, but current brown dwarf models still struggle to fit spectra around the L/T transition with cloud treatment as a major uncertainty \citep{miles_jwst_2023, luhman_jwstnirspec_2023, manjavacas_medium-resolution_2024}. 

Transiting hot Jupiters typically have much lower surface gravity than brown dwarfs and are tidally locked with a permanent nightside. However, the dissipation of MgSiO$_3$ clouds around 1300 K seems to be common between the two classes of objects, potentially hinting at common silicate cloud dissipation mechanisms. 

\subsubsection{Downwell flow or Evaporation?}

We discussed two primary mechanisms that could explain the observed muted morning and clearer evening limbs. While both downwelling flow and dayside evaporation depend on temperature, they are expected to scale differently. Downwelling flow is linked to vertical wind speed, which is driven by the day–night temperature gradient. As a result, hotter planets are expected to exhibit stronger downwelling, leading to a rather monotonic increase in limb asymmetry with temperature (Figure \ref{Fig10:vertical_timescale}). In contrast, evaporation is governed by the crossing of condensation curves, resulting in a more sawtooth-like, non-monotonic relationship with temperature (Figure \ref{Fig8:1D_grid}). If one of the two processes is the dominant driver of inhomogeneous aerosol between the two limbs, more HJs with precise limb spectra across a wide temperature range will be the key to observationally differentiate them.

\begin{table*}[t]
    \centering
    \begin{tabular}{c|c|c|c|c|c|c|c} 
        \hline\hline  
        \bfseries Planet 
        & \bfseries T$_{\text{eq}}$ \rule{0pt}{12pt} 
        & \bfseries Radius 
        & \bfseries Mass 
        & \bfseries log g 
        & \bfseries Morning 
        & \bfseries Evening 
        & \bfseries Eve. - Morn. \\  
        & {\footnotesize (K)} & {\footnotesize ($R_J$)} & {\footnotesize ($M_J$)} & {\footnotesize (cgs)} & (H)  & (H) & (H) \\ 
        \hline
        WASP 107 b	&	770	&	0.94	&	0.10	&	2.45	&	$0.27	\pm	0.09$	&	$0.28	\pm	0.10$	&	$0.02	\pm	0.13$	\\
        WASP 80 b	&	825	&	0.95	&	0.54	&	3.17	&	$0.03	\pm	0.34$	&	$0.68	\pm	0.36$	&	$0.64	\pm	0.49$	\\
        HAT-P-18 b	&	848	&	1.00	&	0.20	&	2.69	&	-$0.11	\pm	0.32$	&	$0.16	\pm	0.32$	&	$0.27	\pm	0.45$	\\
        WASP 39 b	&	1166	&	1.28	&	0.28	&	2.63	&	-$0.21	\pm	0.22$	&	$1.99	\pm	0.22$	&	$2.20	\pm	0.31$	\\
        WASP 96 b	&	1286	&	1.20	&	0.48	&	2.92	&	$0.25	\pm	1.02$	&	$1.15	\pm	1.07$	&	$0.91	\pm	1.48$	\\
        WASP 166 b	&	1272	&	0.63	&	0.10	&	2.8	&	$1.28	\pm	0.90$	&	$1.08	\pm	0.88$	&	-$0.20	\pm	1.26$	\\
        WASP 52 b	&	1301	&	1.27	&	0.46	&	2.85	&	$0.40	\pm	0.29$	&	-$0.22	\pm	0.29$	&	-$0.62	\pm	0.41$	\\
        WASP 94 Ab	&	1543	&	1.58	&	0.50	&	2.7	&	-$0.53	\pm	0.24$	&	$2.30	\pm	0.24$	&	$2.83	\pm	0.34$	\\
        WASP 17 b	&	1740	&	1.99	&	0.51	&	2.51	&	$0.33	\pm	0.17$	&	$1.83	\pm	0.17$	&	$1.50	\pm	0.25$	\\
        \hline
    \end{tabular}
    \caption{Planet parameters and calculated morning and evening A$_H$ values.}
    \label{tab:targets}
\end{table*}

\subsubsection{Cloud or haze?}

The general observational trend of muted morning limbs and clear evening limbs among hotter $\sim>1000K$ planets favors the presence of condensate clouds more than photochemical haze on the morning limb. Since these planets are all tidally locked, photochemically generated hazes are likely to originate on the planet's dayside, which will then be carried by the eastward jet to the planet's evening limb \citep{kempton_observational_2017}. The relatively clearer planet evening limbs disfavor this mechanism. However, with more model complexity, very small ($\sim$nm) haze particles can still be transported to the morning limb \citep{steinrueck_3d_2021}.

Among the cooler $\sim$800 K planets, we observed muted features among both morning and evening limbs on WASP 107 b, WASP 80 b, and HAT-P-18 b. Multiple mechanisms could explain the uniformly muted limbs: (1) The dayside is too cold to evaporate condensate clouds. (2) Dayside haze formation becomes favored and mutes the evening limb. (3) Lower vertical wind speed can no longer effectively remove aerosols from high altitudes. 

To differentiate these mechanisms and definitively determine if the aerosols are cloud or haze, we need broad wavelength coverage of the limb spectra from the UV to mid-infrared. The UV/optical spectrum can constrain the particle size and pressure levels \citep{sing_hststis_2008}. The mid-infrared can spectroscopically resolve the silicate features around 8-10 $\mu m$, enabling the identification of condensate silicate cloud species versus photochemically generated hazes \citep{cushing_spitzer_2006}.

\subsubsection{Caveats}

The simplified timescale calculations and parameterized 1D modeling efforts in this work help us to identify the main physical and chemical drivers for the observed differences in limb spectra across the wide $\sim$1000K temperature range. Compared to more complex GCM and microphysics cloud models, our 1D atmospheric model has a few main caveats and limitations. We assumed the same TP profile shapes, equilibrium chemical composition, Kzz, and cloud condensation lines for all planets. These parameters are likely to be different among the planet samples and can bias the condensation/evaporation interpretation in various ways \citep{zamyatina_observability_2022}. The treatment of clouds is fully parameterized to best fit the data and does not fully capture their underlying formation mechanisms from nucleation to coagulation \citep{powell_transit_2019, kiefer_fully_2024}. We also did not include photochemical haze production \citep{steinrueck_effect_2019}. The grid and data comparison also focuses on the high SNR 1.4$\mu$m water feature rather than fitting the full SOSS spectra. The longer wavelength 1.8$\mu$m and 2.5$\mu$m water features could be affected differently depending on the aerosol optical properties, which are not fully captured by the simplified cloud treatments in the forward model grid. Despite these caveats, our 1D model grid fit serves as a useful illustration of how cloud condensation/evaporation could be a dominant process in driving observed limb spectral differences.

The vertical flow timescale and distance estimates are based on analytical approximations \citep{komacek_vertical_2019}. It demonstrates how downwell flow can effectively remove aerosols on the advection timescales and contribute to observed limb asymmetries. 

To fully explore the effects of all processes at play here, we would need GCM models with fully coupled self-consistent aerosol treatments, which are beyond the scope of this paper.


\section{Biasing atmospheric composition inference}

Typically, atmospheric retrievals assume a uniform limb-averaged spectrum. Way back in 2016, \cite{line_influence_2016} pointed out how nonuniform cloud cover can mimic high mean molecular weight atmospheres and bias retrieval results. With the limited precision from HST WFC3 at the time, we could not observationally confirm this effect. However, with JWST, we show definitive evidence for highly nonuniform cloud coverages between the morning and evening limbs for at least three out of the nine hot Jupiters. Analytically, the wavelength-dependent transit spectra with patchy clouds can be approximated with equation 10 from \citep{line_influence_2016}:

\begin{equation}
\frac{d\alpha_\lambda}{d\lambda} = (1 - f) \frac{2R_p}{R_{star}^2} H \frac{d \ln(\sigma_{\lambda, \mathrm{H_2O}})}{d\lambda}
\end{equation}

Parameter $f$, the terminator cloud fraction, is degenerate with the scale height ($H$), which determines all atmospheric molecular feature sizes. There are two main knobs we turn in the scale height equation when performing atmospheric properties inference, mean molecular weight $\mu$, and temperature. If $\frac{d\alpha_\lambda}{d\lambda}$ gets reduced by half from averaging the featureless morning limb with molecular features in the evening limb, the true $f$ should be 1/2. However, when it is neglected (equivalent to f=0), $H$ will then be reduced by half to compensate. Thus, $\mu$ and temperature become biased. Since temperature is the numerator and $\mu$ is the denominator of the scale height equation, they suffer from similar but inverse biases. Specifically, $\mu$ will be biased towards higher values while temperature will be biased towards lower values. 

\begin{figure}[htp]
    \centering
    \includegraphics[width=0.45\textwidth]{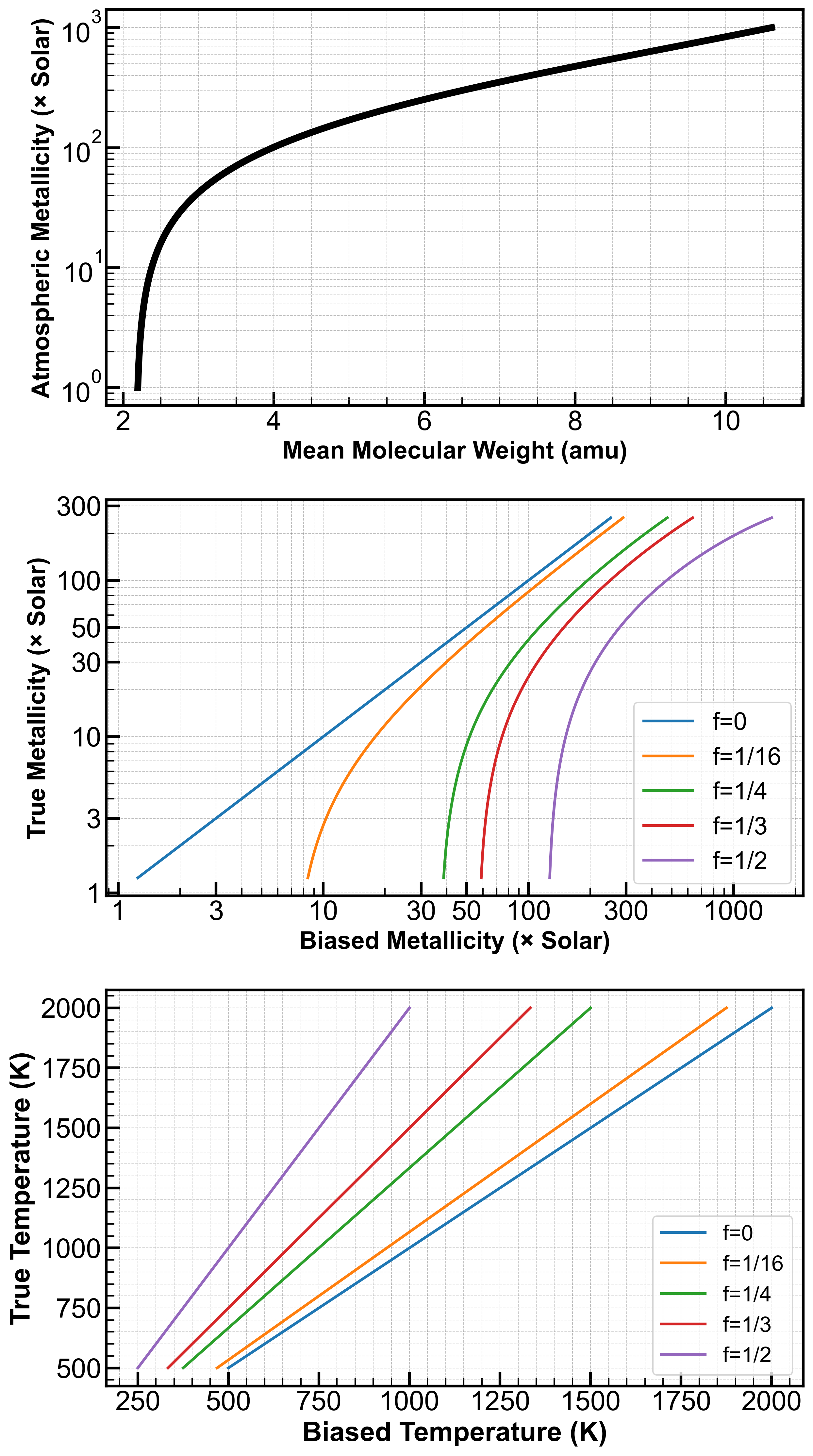}
    \caption{The muted morning and clear evening limbs can mimic a higher metallicity and/or cooler uniform limb. Based on equation 10 from \citep{line_influence_2016}, we calculated the biased metallicity and temperature as functions of their respective true values in 5 cases. In the limit where the morning limb is completely featureless and the evening limb is clear (f=1/2), metallicity can be biased upward by $\sim$2 dex, or temperature can be biased downward by half.}
\label{Fig11:bias}   
\end{figure}

To demonstrate this, we first calculated the mean molecular weight as a function of atmospheric metallicity assuming solar composition (Top panel, Figure \ref{Fig11:bias}). Next, based on a range of $\mu$ values, we calculated the true metallicity and biased metallicity by multiplying the $\mu$ values by 1/(1-$f$). For a planet with a true abundance of $10\times$ enhanced metallicity (Z), its inferred metallicity under uniform limbs assumption could be biased upward by $\sim2$~dex if it has featureless morning and clear evening limbs. Next, we show the opposite biasing effect on temperature. We plotted a range of true temperatures and biased temperatures with a multiplier of $1-f$ in the bottom panel of Figure \ref{Fig11:bias}. Depending on the $f$ factor, the true planet temperature could be biased to lower temperatures by up to half.

Compared to limb temperatures, metallicity is often the parameter of greater interest in atmospheric studies, and this information is mainly embedded in two ways in the observed transit spectra: (1) the absolute amplitude of all molecular features. (2) the relative amplitudes between different molecules (e.g., H$_2$O vs CO$_2$). While the former relies on measuring the absolute elemental abundances, the latter leverages elemental ratios information from chemistry. (1) is more important in free retrieval frameworks, and (2) is a stronger constraint in grid retrievals, often assuming equilibrium chemistry. When $H$ is reduced, it affects (1) more than (2). Therefore, if the goal is to accurately measure metallicity, grid retrievals with the dataset covering H$_2$O vs CO$_2$ are recommended.

Among the three planets showing such strong limb asymmetries, we can see these effects in the reported metallicity and temperatures. The WASP-94 Ab study with NIRSpec G395H reported $\sim$2x solar metallicity \citep{ahrer_tracing_2025}, consistent with the SOSS study, which retrieves the two limbs separately \citep{mukherjee_cloudy_2025}. However, \cite{ahrer_tracing_2025} reported lower temperature ranges ($\sim$700-1000K) compared to $\sim$1500K from \cite{mukherjee_cloudy_2025}. Given the $T_{eq}\sim$1550K of WASP-94 Ab, temperatures retrieved in \cite{ahrer_tracing_2025} are likely biased by reduced $H$ from unaccounted limb asymmetries. Despite the biased lower temperature, the relative H$_2$O to CO$_2$ features ensured the metallicity from the grid retrieval with the equilibrium chemistry assumption remains largely robust. However, the free retrieval struggles more in comparison. The reported log$_{10}$ H$_2$O=-1.60 \citep{ahrer_tracing_2025} corresponds to $\sim$40x solar H$_2$O abundance. The study of WASP-17 b SOSS transit spectrum reported a retrieved temperature of 1272 K, which is also low relative to its equilibrium temperature of 1755 K \citep{louie_jwst-tst_2024}. Retrievals of WASP-39 b also suffer the similar issues, with SOSS being the most affected due to relatively stronger muting effects from aerosols. In \cite{lueber_information_2024}, the reported water abundance (log$_{10}$ H$_2$O=-1.13) is $\sim$2 dex higher than the PRISM (log$_{10}$ H$_2$O=-3.1), and $\sim$1 dex higher in metallicity than reported in \cite{espinoza_inhomogeneous_2024}, which used grid retrieval to fit the two limbs separately.

Multiple studies have explored the biasing effects of day-night temperature gradients \citep{caldas_effects_2019, pluriel_strong_2020} and aerosols heterogeneous \citep{lacy_jwst_2020, welbanks_atmospheric_2022} on retrieving limb-averaged spectra with 1D models. However, it is yet to be demonstrated that more complex retrieval setups can fully account for these biasing effects \citep{mukherjee_cloudy_2025, chen_asymmetry_2025}. We have shown that aerosol heterogeneity is common but not universal in hot Jupiters, and how they can directly reduce the scale heights in limb-averaged spectra and bias temperature and/or metallicity inferences in existing literature. Given that we can now directly resolve the two limb spectra, further studies to benchmark the robustness of retrieving limb-averaged versus limb-resolved spectra are needed.

Most transiting exoplanets, including sub-Neptunes and terrestrial planets, have short orbital periods and are expected to be tidally locked. Clouds are also expected to form in low-mass planets with temperate atmospheres from Trappist-1 e to K2-18 b \citep{charnay_formation_2021, barrier_new_2025}. A similar mechanism of different temperatures on the two terminators straddling cloud condensation lines and causing cloudy evenings and clear evenings can also take place. The expected prevalence of photochemical haze \citep{kempton_reflective_2023, gao_hazy_2023} on the dayside can further complicate the picture by muting evening limbs as well.

\section{Limb Spectroscopy Metric (LSM)}

To quantify the limb-limb asymmetry signal size and guide future observations, we developed the Limb Spectroscopy Metric (LSM). It is similar to TSM \citep{kempton_framework_2018} and Exo.MAST observability SNR metric \citep{mullally_exomast_2019}, but takes ingress/egress and impact parameters into account. We first calculated the ingress/egress duration $\tau$ using the following equation from \citep{espinoza_constraining_2021}:

\begin{equation}
\tau = \left( \frac{P}{\pi} \right) \left( \frac{1}{\sqrt{1 - b^2}} \right) \left( \frac{R_p}{R_*} \right) \left( \frac{R_*}{a} \right)
\end{equation}

The parameters are defined in the same way as in \cite{espinoza_constraining_2021} with $P$ as planet orbital period, $b$ as the impact parameter, $R_p$ as planet radius, $R_*$ as stellar radius and a as semi-major axis.

Next, we define the one-scale-height spectral feature size $\delta$ the same way as in \citep{mullally_exomast_2019}:

\begin{equation}
\delta = \frac{2 H R_p}{R_*^2}
\end{equation}

The atmospheric scale height ($H$) is defined as the following:
\begin{equation}
H = \frac{K T_{\text{eq}}}{\mu g}
\end{equation}

where $K$ is the Boltzmann constant, $T_{eq}$ is the planet equilibrium temperature, $\mu$ is the mean molecular weight assuming hydrogen-dominated atmosphere ($\mu$=2.3 amu), and g is the planetary surface gravity. 

\begin{equation}
\text{LSM} = \left( \frac{\delta}{\delta_{\text{ref}}} \right) \sqrt{10^{-0.4(J - J_{\text{ref}})}} \sqrt{\frac{\tau}{\tau_{\text{ref}}}} 
\end{equation}

The LSM is defined using WASP 94 Ab as the reference planet with $\delta_{ref}$=234.6ppm, $J_{ref}$=9.159 mag, and $\tau_{ref}$=25.22 minutes. They are calculated based on the following parameters: $R_p$=1.58R$_J$, $R_*$=1.48R$_s$, $T_{eq}$=1543K, $M_p$=0.5M$_J$, $g=4.96$m/s$^2$, $P=3.95019$~days and $a=0.055$~au.

We also added two terms to account for multiple number of transit visits ($N$) and different observatories with different aperture diameters ($D$). The two terms are simple scaling assuming Poisson photo noise. 

\begin{equation}
\text{LSM}_{N, telescope} \propto {\sqrt N} \frac{D}{6.5m}
\end{equation}

\begin{figure}[htp]
    \centering
    \includegraphics[width=0.48\textwidth]{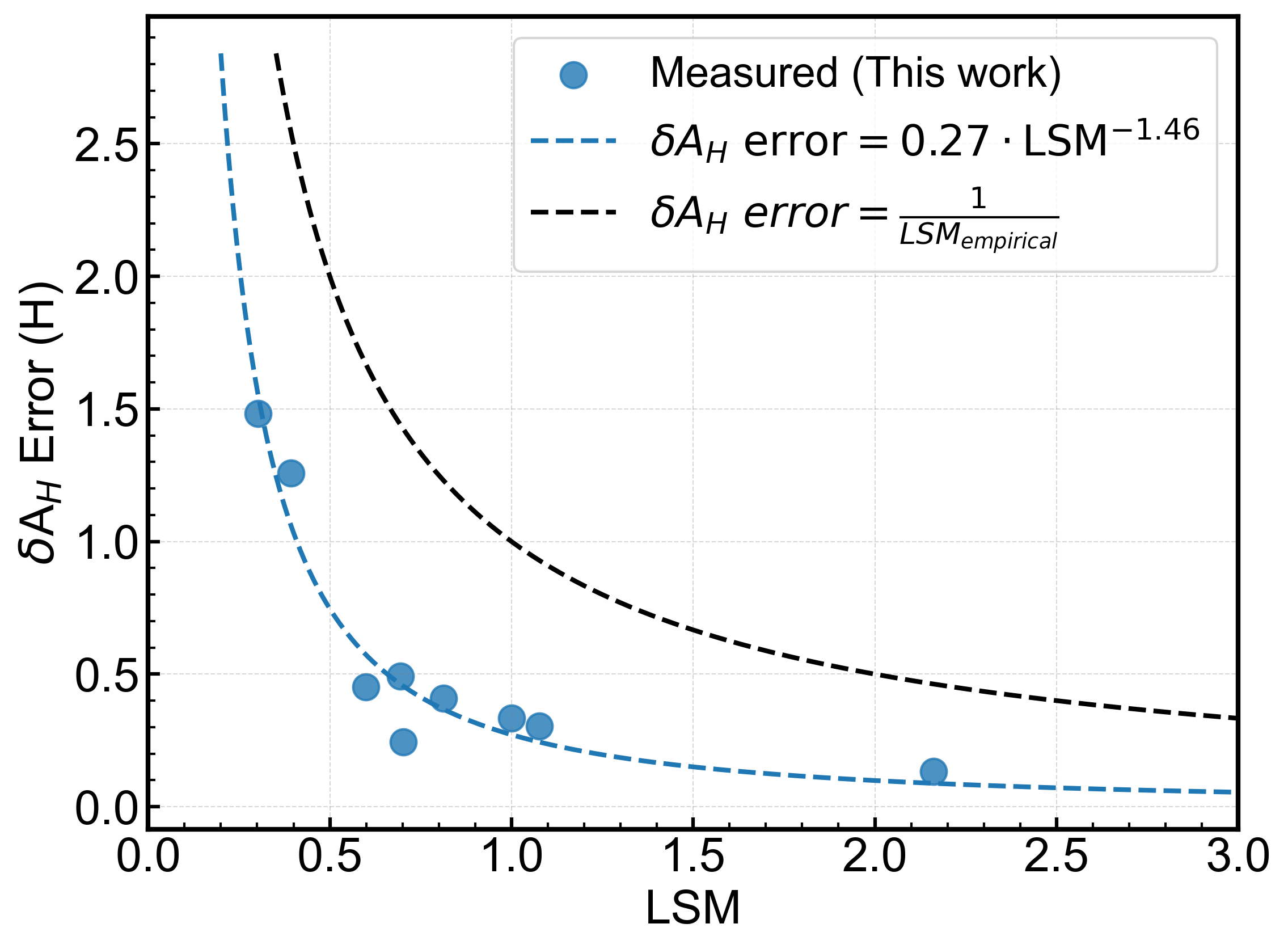}
    \caption{We fitted a power law function to the measured $\delta$A$_H$ error and calculated LSM values based on the nine planets in this work. Then we defined LSM$_{empirical}$ so it scales inversely to $\delta$A$_H$ error.}
\label{Fig12:LSM_fit}   
\end{figure}

For the LSM to better reflect expected JWST observation, we measured LSM for all SOSS datasets and fitted a power law to LSM versus measured $\delta$A$_H$ error bar ($\delta$A$_H$err) sizes (Figure \ref{Fig12:LSM_fit}). Based on this empirically determined relationship, we rescale LSM to LSM$_{empirical}$: 

\begin{equation}
\text{LSM$_{empirical}$} = \frac{1}{0.27 * LSM^{-1.46}}
\end{equation}

LSM$_{empirical}$ = 1 would then correspond to a measured $\delta$A$_H$err of 1H and LSM$_{empirical}$ is inverse to $\delta$A$_H$err:

\begin{equation}
\text{dA}_H\ \text{err (H)} = \frac{1}{\text{LSM}_{\text{empirical}}} 
\end{equation}

\begin{figure}[htp]
    \centering
    \includegraphics[width=0.45\textwidth]{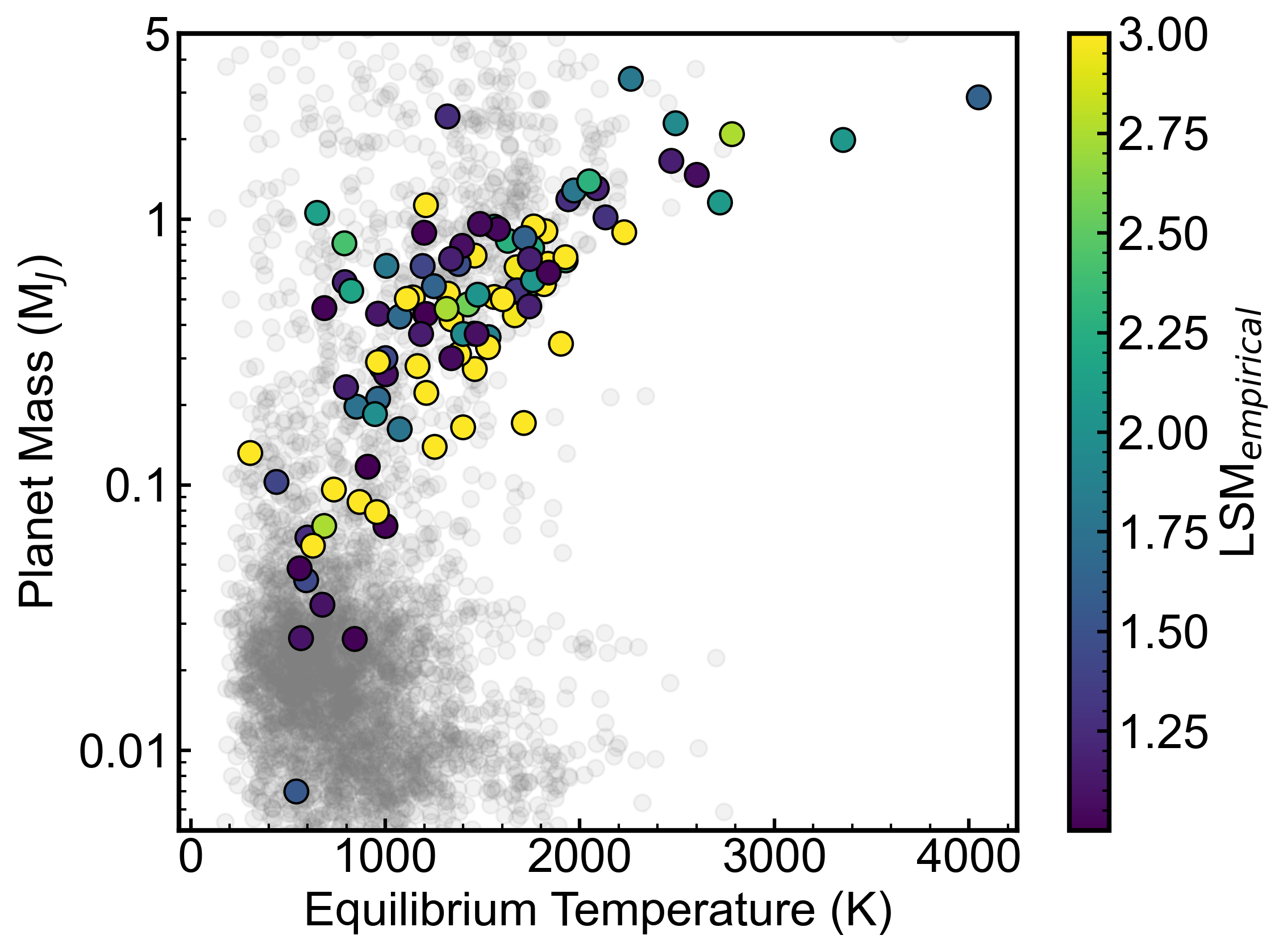}
    \caption{All known transiting exoplanets (grey) to date and the 105 planets (color) with LSM$_{empirical}>$1.}
\label{Fig13:LSM}   
\end{figure}

LSM$_{empirical}$ is an empirically-derived index based on the JWST SOSS datasets and data analysis routines described in this paper. It scales inversely to $\delta$A$_H$err. It serves as an approximate guiding metric for future JWST observations. A planet with LSM$_{empirical}$=1 is expected to have the precision to distinguish between the morning and evening limb spectral shape differences in and out of the 1.4$\mu$m water to 1 atmospheric scale height ($H$) precision. There are $\sim$105 targets from the NASA exoplanet archive to date with LSM$_{empirical}>$1 (Figure \ref{Fig13:LSM}).

\section{Conclusion}

We present the limb spectra for nine transiting planets observed with NIRISS/SOSS. Among the nine planets, three show prominent limb-to-limb spectral differences. To quantify the observed limb spectra at a population level, we adopted a limb water index (A$_H$) measuring the relative in versus out of the 1.4 $\mu$m water band transit depth normalized in atmospheric scale heights. Then we correlated them with planetary equilibrium temperature and surface gravity. Based on the observed A$_H$ and $\delta$A$_H$, we have the following key findings:

(1) Observed muted morning limbs and clear evening limbs require atmospheric processes that can remove high-altitude aerosols on a timescale comparable to the advection timescale ($\sim$day). Both the downwelling of the vertical flow and dayside evaporation could be plausible mechanisms. Gravitational settling can remove large grains, but is ineffective for smaller particles, even with added stellar radiation pressure. 

(2) The simplest explanation for the morning aerosol composition is nightside condensate clouds. Dayside photochemically generated haze is expected to be circulated to the evening limb first, and the relatively clear evening limbs disfavor a significant presence of dayside haze. 

(3) While morning limbs stay muted for all planets, the evening limbs clear up at around 1200 K and 1600 K, forming a double peak feature versus planetary equilibrium temperature with a valley at around 1300 K. By comparing to 1D \texttt{picaso} forward model grid, we found that a single cloud species like MgSiO$_3$ can only produce one morning-evening limbs difference at around 1600 K, and a second cloud population (Na$_2$S and MnS) is needed to produce the other limb-limb spectra difference in WASP 39 b. The MgSiO$_3$ needs to dissipate at around 1300 K for the evening limb of WASP 39 b to clear up. This is analogous to the L/T transition in brown dwarf atmospheres, where silicate cloud features disappear in the spectra faster than model predictions. 

(4) We hypothesize an empirically derived transition line, "asymmetry horizon", which marks where planets are expected to start exhibiting muted morning and clearer evening limbs from aerosols heterogeneity. This transition line goes across in temperature and surface gravity parameter space. It will be easily testable with future observations.

(5) The inhomogeneous limb aerosol coverage, as shown by large $\delta$A$_H$ values, can mimic limb-averaged spectra with smaller scale heights. If left uncorrected, it can bias retrieval towards higher inferred metallicity or lower temperature. The common large $\delta$A$_H$ in the nine-planet sample requires future hot Jupiter atmospheric composition measurements to take this effect into account. 

(6) The Limb Spectroscopy Metric (LSM) is defined to better estimate the expected signal-to-noise ratio of the limb spectra by taking ingress/egress duration and impact parameter into account. We also empirically derived LSM$_{empirical}$ based on the measured $\delta$A$_H$ error bar sizes from SOSS datasets, so it scales more intuitively to the observation.

Ground-based high-resolution spectroscopy has been sensitive to the presence of individual chemical species between the morning and evening limbs \citep{ehrenreich_nightside_2020, kesseli_atomic_2022}, but the interpretations are often challenging due to degeneracies between chemical gradients or clouds \citep{savel_diagnosing_2023}. We show here that JWST has the unique capability to definitively spectroscopically detect the presence of clouds on the limbs. This opens up new synergies with ground-based observations to better constrain the morning-to-evening limb atmospheric chemistry difference.  

All of these findings are falsifiable and will be tested as more planets are observed by JWST with published morning/evening limb spectra. The implications for smaller planets, such as sub-Neptunes and rocky planets, are currently unknown. Such planets are usually cooler ($<$1000 K) but still expected to be tidally locked. Similar mechanisms of morning and evening limbs straddling cloud condensation lines lead to cloudy mornings and clear evenings can also take place. However, it is challenging to detect their limb-to-limb asymmetry signals due to their much smaller atmospheric feature sizes.

\vspace{1cm}
G.F. thanks Thaddeus Komacek for the insightful discussions on the ski slopes during the conference at the Aspen Center for Physics. We thank the anonymous referee for the thorough and useful comments and suggestions. 

This work is based on observations made with the NASA/ESA/CSA James Webb Space Telescope. The data presented in this article were obtained from the Mikulski Archive for Space Telescopes (MAST) at the Space Telescope Science Institute, which is operated by the Association of Universities for Research in Astronomy, Inc., under NASA contract NAS 5-03127 for JWST. These observations are associated with program \#5924. Support for program \#5924 was provided by NASA through a grant from the Space Telescope Science Institute, which is operated by the Association of Universities for Research in Astronomy, Inc., under NASA contract NAS 5-03127. The specific observations analyzed can be accessed via \dataset[DOI: 10.17909/j0vy-by23]{https://doi.org/10.17909/j0vy-by23}.

Spectra from Figure 2 are available to download via \dataset[DOI:10.5281/zenodo.16276983]{https://doi.org/10.5281/zenodo.16276983}.

\section{Appendix A \\ \texttt{picaso} 1D forward model grid}

\begin{figure}[htp]
    \centering
    \includegraphics[width=0.45\textwidth]{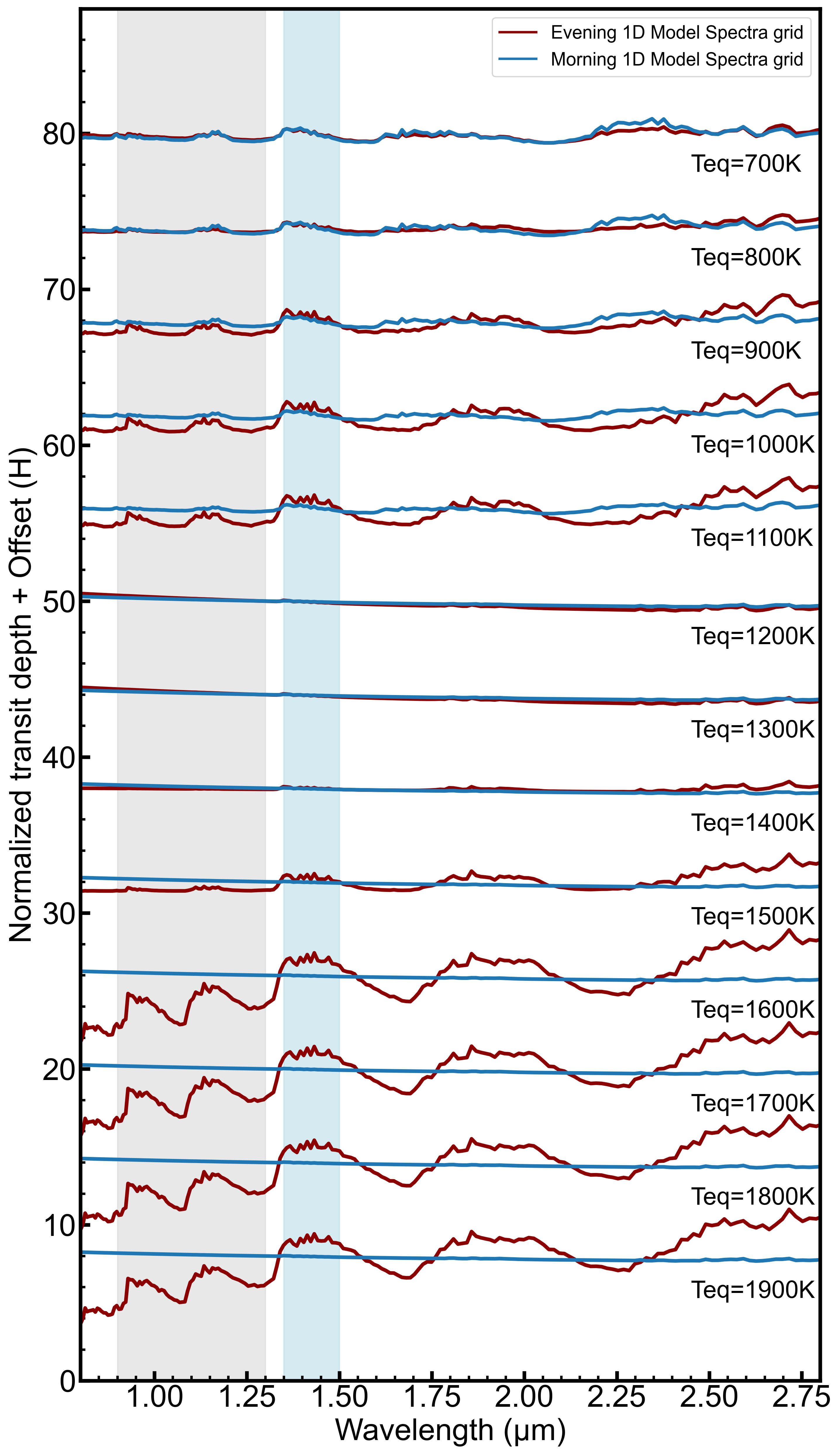}
    \caption{The scale height normalized 1D forward model grid for morning and evening limbs as shown in the top left panel of Figure \ref{Fig8:1D_grid}. The plotted spectra sample the temperature range by every 100 K from 700 to 1900K.}
\label{Fig14:appendix_model_grid}   
\end{figure}

To interpret the observed A$_H$ between the morning and evening limbs across different temperature ranges, we generated 1D forward models with \texttt{picaso} as described in the section 2.5 and shown in Figure \ref{Fig8:1D_grid}. Here we show the forward models (Figure \ref{Fig14:appendix_model_grid}) used to create the model tracks in Figure \ref{Fig8:1D_grid} for both limbs sampled at every 100K. Both limb model spectra are normalized by the same scale height, assuming equilibrium temperatures to ensure consistent comparison with the observed spectra, which are also normalized by the planet's equilibrium temperatures.

\end{document}